\documentclass[a4paper,onecolumn,11pt,accepted=2026-10-12]{quantumarticle}
\pdfoutput=1

\usepackage[utf8]{inputenc}
\usepackage[english]{babel}
\usepackage[T1]{fontenc}
\usepackage{lipsum}
\usepackage{soul}
\soulregister\cite7
\soulregister\citep7
\soulregister\eqref7
\soulregister\ref7
\renewcommand{\hl}[1]{#1}

\usepackage{graphicx}
\usepackage{dcolumn}
\usepackage{bm}

\usepackage[margin=1in]{geometry}
\usepackage{amsmath,amssymb,amsthm,amsfonts}
\usepackage{array}
\usepackage{caption}
\usepackage{ragged2e}
\usepackage{subcaption}
\usepackage{booktabs}
\usepackage{multirow}
\usepackage{hyperref}
\usepackage{color}
\usepackage{braket}
\usepackage{nicematrix}
\usepackage{tikz}
\usepackage{graphicx}
\usetikzlibrary{fit}
\usepackage{scalerel}

\usepackage[most]{tcolorbox}
\usepackage{xcolor}
\newtcolorbox{highlightblock}{
  enhanced,
  breakable,
  colback=yellow!25,
  colframe=yellow!25,
  boxrule=0pt,
  arc=0pt,
  outer arc=0pt,
  left=2pt,
  right=2pt,
  top=2pt,
  bottom=2pt,
  before skip=0.5\baselineskip,
  after skip=0.5\baselineskip,
  parbox=false
}
\renewenvironment{highlightblock}{}{} 

\DeclareMathOperator*{\concat}{\scalerel*{\Vert}{\sum}}

\newtheorem{theorem}{Theorem}

\newtheorem{observation}{Observation}
\newtheorem{remark}{Remark}

\newcommand{\I}{\mathbb{I}}
\newcommand{\Tr}{\mathrm{tr}}

\newcommand{\ketbraauto}[1]{\left|#1\right\rangle\left\langle#1\right|}
\newcommand{\braauto}[1]{\left\langle#1\right|}
\newcommand{\ketauto}[1]{\left|#1\right\rangle}
\newcommand{\R}{\mathbb{R}}
\newcommand{\norm}[1]{\left\lVert #1 \right\rVert}

\begin{document}

\title{Re-uploading quantum data: a universal function approximator for quantum inputs}%

\author{Hyunho Cha}
\email{ovalavo@snu.ac.kr}
\affiliation{NextQuantum and Department of Electrical and Computer Engineering, Seoul National University, Seoul 08826, Republic of Korea}
\orcid{0009-0008-2933-6989}
\author{Daniel K. Park}
\email{dkd.park@yonsei.ac.kr}
\orcid{0000-0002-3177-4143}
\affiliation{Department of Statistics and Data Science, Yonsei University, Seoul 03722, Republic of Korea}
\affiliation{Department of Applied Statistics, Yonsei University, Seoul 03722, Republic of Korea}
\affiliation{Department of Quantum Information, Yonsei University, Incheon 21983, Republic of Korea}
\author{Jungwoo Lee}
\email{junglee@snu.ac.kr}
\affiliation{NextQuantum and Department of Electrical and Computer Engineering, Seoul National University, Seoul 08826, Republic of Korea}
\orcid{0000-0002-6804-980X}
\maketitle

\begin{abstract}
Quantum data re-uploading has proved powerful for classical inputs, where repeatedly encoding features into a small circuit yields universal function approximation. Extending this idea to quantum inputs remains underexplored, as the information contained in a quantum state is not directly accessible in classical form.
We propose and analyze a quantum data re-uploading architecture in which a qubit interacts sequentially with fresh copies of an arbitrary input state. \hl{The circuit can approximate any bounded continuous function using only one ancilla qubit and single-qubit measurements, provided that sufficiently many copies of the input state and, in general, sufficiently many layers are available.} By alternating entangling unitaries with mid-circuit resets of the input register, the architecture realizes a discrete cascade of completely positive and trace-preserving maps, analogous to collision models in open quantum system dynamics. Our framework provides a \hl{space-efficient} and expressive approach to designing quantum machine learning models that operate directly on quantum data.
\end{abstract}

\section{Introduction}

Quantum machine learning (QML) seeks to harness quantum computation to enhance machine learning tasks \cite{schuld2015introduction, biamonte2017quantum, moll2018quantum, huang2021power, liu2021rigorous}. Quantum computers can perform certain linear algebra subroutines faster than classical machines under state preparation assumptions \cite{harrow2009quantum, lloyd2013quantum, lloyd2014quantum}. Motivated by such potential quantum speedups, a variety of QML models have been explored to outperform their classical counterparts, including quantum kernel methods and variational quantum circuits \cite{benedetti2019parameterized, schuld2019evaluating, schuld2019quantum, sim2019expressibility, blank2020quantum, cerezo2021variational, schuld2021quantum, jerbi2023quantum}.

A key component of any QML model is how data are encoded into and processed by quantum circuits \cite{caro2021encoding, schuld2021effect, schuld2021supervised, li2022concentration, daimon2024quantum, gonzalez2024efficient, hur2024neural, jaderberg2024let, rath2024quantum, liu2025neural}. For classical input data, one common approach is to embed the data into a quantum state through parameterized gate operations. Recent work has shown that repeatedly encoding data within a circuit, known as data re-uploading, enhances a model's expressive power and can even make a single qubit a universal quantum classifier \cite{perez2020data, perez2021one,schuld2021effect, PhysRevA.111.022429}. The key idea is to interleave data-encoding gates with trainable quantum gates across several layers, effectively re-uploading the input at each layer. Similarly, quantum signal processing techniques were employed to prove that a sequence of single-qubit rotations interleaved with data-dependent operations can implement polynomial transformations of a classical input \cite{low2017optimal, martyn2021grand, rossi2022multivariable, motlagh2024generalized}.

Such results underscore that parameterizing quantum gates with classical data is a powerful paradigm in QML, enabling even small circuits to represent complex functions of the input data.
Nonetheless, recent work has questioned whether quantum models trained on purely classical datasets can demonstrate genuine quantum advantage \cite{kubler2021inductive}. On the other hand, quantum advantage is more naturally expected when models are applied to quantum data \cite{cerezo2022challenges} with concrete demonstrations emerging in recent studies \cite{huang2022quantum}.
However, when the input data are quantum rather than classical, this direct re-uploading approach encounters a fundamental obstacle. In a quantum input setting (where data are provided as quantum states), one cannot simply plug the input state's ``values'' into rotation angles or other gate parameters, because those values are not explicitly available as classical numbers. Thus, a new strategy is required to re-upload quantum data in a multi-layer quantum model.

\hl{
A closely related line of work studies direct multicopy estimation of nonlinear functionals of quantum states. Controlled-SWAP and cyclic-shift interferometric networks can estimate overlaps, purities, and spectral moments such as $\Tr(\rho^k)$ without first reconstructing $\rho$ \cite{ekert2002direct}. More generally, any $m$th-degree polynomial in the density matrix entries can be represented as the expectation value of an observable on $m$ copies of $\rho$ \cite{brun2004measuring}. Recent quantum state function methods make this viewpoint sample-efficient by combining generalized SWAP tests with linear combinations of unitaries and parameterized circuits, enabling normalized degree-$p$ polynomial functionals to be estimated to precision $\varepsilon$ using $O(p/\varepsilon^2)$ copies, with applications to entropy, fidelity, and eigenvalue estimation \cite{yao2026sample}. Our goal is complementary: rather than designing a dedicated multicopy measurement for a specified functional, we analyze a general trainable sequential re-uploading architecture in which a single qubit coherently interacts with fresh copies of the input, yielding an ansatz with universal approximation guarantees for continuous functions of quantum states.
}

More specifically, we introduce a \emph{quantum data re-uploading} architecture. The proposed model consists of a single-qubit \emph{signal} register that interacts with the quantum input state in a sequence of layers. Instead of encoding quantum data as a gate parameter, each layer involves entangling the signal qubit with the input state, thereby uploading the input's information into the joint system. In essence, the signal qubit acts as an intermediary that undergoes quantum operations that depend on both the input and learned parameters. By repeating this step across multiple layers, the circuit incrementally builds a complex transformation of the input state.

We show that this re-uploading architecture is capable of universal function approximation on quantum data while using only a single qubit for the processing unit. In fact, no matter how many qubits the input state contains, the model requires just one extra qubit (the signal register) to achieve universality. This result extends the spirit of single-qubit universality to the domain of quantum inputs.

A key feature of the proposed approach is its efficient use of qubits through qubit reuse \cite{botelho2022error, brandhofer2023optimal, decross2023qubit, zhu2023interactive}. After the signal qubit interacts with the input in each layer, we simply discard the second register and then reset it to the input state. The signal register is left unmeasured so that it can continue to the subsequent layer. This qubit-reuse tactic allows the same register to load data in all layers, rather than needing a fresh input register for each layer. As a result, the total number of qubits required by the model does not scale with the number of layers, since only the input register and one working qubit are needed. This approach is compatible with the capabilities of many modern quantum devices that support reset operations.


\section{Notations}
Let \(n\) be the number of qubits in the input system, and set \(d=2^n\). We denote the associated \(d\)-dimensional Hilbert space by \(\mathcal{H}_d\). The set of all density operators on \(\mathcal{H}_d\) is denoted as
\[
\mathcal{D}(\mathcal{H}_d) = \{\rho\in\mathcal{B}(\mathcal{H}_d)\mid\rho=\rho^\dagger,\,\rho\succeq0,\, \Tr(\rho)=1\},
\]
where \(\mathcal{B}(\mathcal{H}_d)\) denotes the space of all bounded linear operators on \(\mathcal{H}_d\). The corresponding space of observables is the vector space of Hermitian operators, which we denote by
\[
\operatorname{Herm}(\mathcal{H}_d) = \{O\in\mathcal{B}(\mathcal{H}_d)\mid O=O^\dagger\}.
\]
The identity operator on \(\mathcal{H}_d\) is denoted by \(\I_d\). We adopt the notation
\[
\left(\sigma^{(0)}, \sigma^{(1)},\sigma^{(2)},\sigma^{(3)}\right) = (\I_2, X,Y,Z)
\]
to refer to the standard Pauli matrices.

\section{Main results}

\begin{figure}
    \centering
    \includegraphics[width=0.9\linewidth]{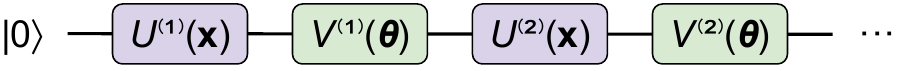}
    \caption{A single-qubit classical data re-uploading scheme. \(U^{(l)}(\mathbf{x})\) denotes an input-dependent data encoding gate, while \(V^{(l)}(\boldsymbol\theta)\) denotes a trainable input-independent gate.}
    \label{fig:reupload_classical}
\end{figure}

Re-uploading classical input data into quantum circuits for function approximation is a well-established technique \cite{perez2020data, martyn2021grand, perez2021one}. In this approach, classical input data \(\mathbf{x}\) is encoded into a quantum circuit multiple times through parameterized gates, which alternate with trainable gates parameterized by \(\boldsymbol\theta\) (see Figure~\ref{fig:reupload_classical}). However, when the input data is quantum, its density matrix elements cannot be used directly to parameterize the circuit,  as a full classical description of the state is generally not efficiently accessible. This naturally leads to the case of directly re-uploading the quantum data.

\begin{figure}
    \centering
    \begin{subfigure}{0.9\linewidth}
        \centering
        \includegraphics[width=\linewidth, page=1]{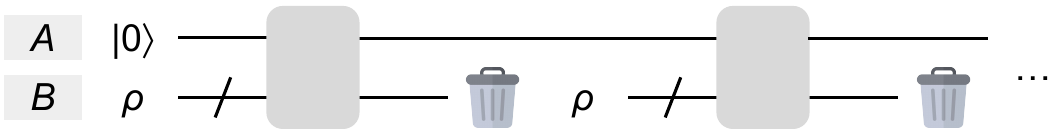}
        \caption{\centering}
        \label{fig:reupload_general}
    \end{subfigure}
    \vspace{1em}
    \begin{subfigure}{0.86\linewidth}
        \centering
        \includegraphics[width=\linewidth, page=1]{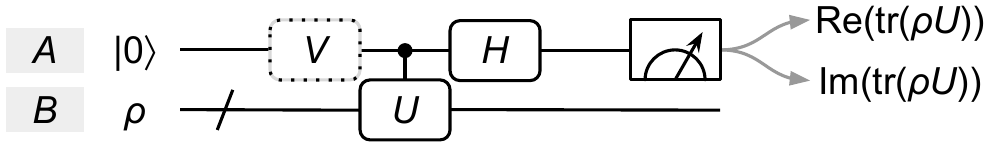}
        \caption{\centering}
        \label{fig:trace_estimation}
    \end{subfigure}
    \caption{(a) A general quantum data re-uploading scheme for universal function approximation. \(\rho\in\mathcal{D}(\mathcal{H}_d)\) denotes the quantum input and gray boxes indicate arbitrary \(\text{SU}(2d)\) elements. (b) The Hadamard test circuit can be used to estimate \(\Tr(\rho U)\) by measuring the expectation value \(\langle Z\rangle\) on system \(A\), using two separate configurations: \(V=H\) to obtain \(\text{Re}(\Tr(\rho U))\), and \(V=S^\dagger H\) to obtain \(\text{Im}(\Tr(\rho U))\).}
\end{figure}

Let \(\rho\in\mathcal{D}(\mathcal{H}_d)\) be input quantum data. Let \(A\) and \(B\) denote the signal and input registers, respectively, where \(\rho\) is loaded into \(B\). The circuit illustrated in Figure~\ref{fig:reupload_general} effectively accumulates information about \(\rho\) into \(A\), which is initialized in the state \(\ket{0}\). After \(L\) re-uploading layers, we obtain a final density matrix on register \(A\). A classical post-processing step is then applied to extract a single real scalar output.

\begin{remark}
\hl{
When $L=1$, the re-uploading model resembles the Hadamard test. The circuit can be used to estimate the value of $\Tr(\rho U)$ for a given unitary $U\in\textnormal{SU}(d)$ (see Figure~\ref{fig:trace_estimation}).
For the maximally mixed input $\rho=\I_d/d$, the estimated quantity reduces to the normalized trace $\Tr(U)/d$, which appears in the trace estimation setting of the one clean qubit model of computation, also known as \textit{deterministic quantum computation with one clean qubit} (DQC1) \cite{knill1998power,shepherd2006computation}. This observation provides an informal connection between the $L=1$ re-uploading circuit and DQC1 trace estimation.
}
\end{remark}

\begin{remark}
\hl{
A standard variational quantum classifier (VQC) that processes only one copy of the quantum input state is contained in the $L=1$ case of our quantum data re-uploading model, at least for the usual setting of a variational unitary followed by a final observable on the first qubit. Consequently, in tasks where the target function requires higher-order dependence on the input state and is achieved only for $L>1$, the re-uploading architecture can represent functions that standard single-copy VQCs of this form cannot.
}
\end{remark}

To realize the circuit shown in Figure~\ref{fig:reupload_general}, we assume the ability to reset system \(B\) to the state \(\rho\). Although qubit reuse remains relatively underexplored in quantum circuit optimization, this capability is gaining increasing attention as it can significantly reduce the total number of qubits required in a computation \cite{brandhofer2023optimal, decross2023qubit}. As discussed in Section~\ref{sec:explicit_model_interpretation}, the circuit in Figure~\ref{fig:reupload_general} is logically equivalent to an \((Ln+1)\)-qubit circuit, where \(L\) is the number of re-uploading layers. However, with qubit reuse, the same computation can be implemented using only \((n+1)\) qubits, independent of \(L\).

The reset of register $B$ to $\rho$ should be understood as fresh-copy access to the input state, rather than a standard hardware reset to a fixed reference state such as $|0\rangle^{\otimes n}$. Consequently, an $L$-layer circuit requires $L$ preparations or copies of $\rho$. The total execution cost therefore depends on the state-preparation channel.

\subsection{Re-uploading a single-parameter single-qubit state}
\label{section:single_parameter_state}

\begin{figure}
    \centering
    \includegraphics[width=0.9\linewidth, page=2]{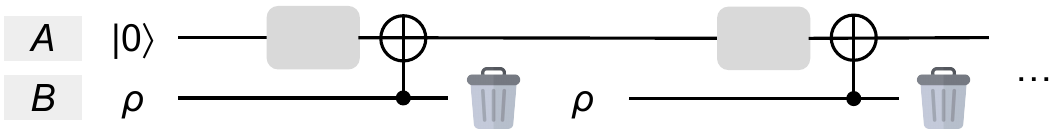}
    \caption{A restricted single-qubit data re-uploading scheme for universal function approximation. \(\rho\in\mathcal{D}(\mathcal{H}_2)\) denotes the single-qubit input and gray boxes indicate arbitrary \(\text{SU}(2)\) elements.}
    \label{fig:reupload_restricted}
\end{figure}

We aim to establish that the re-uploading structure in Figure~\ref{fig:reupload_general} possesses universal function approximation capability. Consider the simplest case where \(\rho\) is a single-parameter single-qubit pure state with real amplitudes:
\[
\rho(t) = |\psi(t)\rangle\langle\psi(t)| \in \mathcal{D}(\mathcal{H}_2),
\]
where
\begin{equation}
\label{equation:psi_t_vector}
\ket{\psi(t)} = t\ket{0}+\sqrt{1-t^2}\ket{1} \in \mathcal{H}_2, \quad 0<t<1.
\end{equation}
Also, let \(\tau^{(l)}\) denote the density matrix of \(A\) after \(l\) re-uploading layers. Without loss of generality, we take \(\tau^{(0)} = |+\rangle\langle+|\) in place of \(|0\rangle\langle0|\). We further simplify the scheme by restricting each \(\text{SU}(4)\) element (a gray box in Figure~\ref{fig:reupload_general}) to consist of a single-qubit gate on \(A\) followed by \(\text{CNOT}_{B\rightarrow A}\) (see Figure~\ref{fig:reupload_restricted}). In this restricted form, the circuit structure mirrors the pattern of interleaving trainable gate layers with fixed, data-encoding operations, which in turn simplifies our subsequent analysis.

Let
\[
\mathbf{r}^{(l)}=\left(r_1^{(l)},r_2^{(l)},r_3^{(l)}\right)=\left(\left\langle\sigma^{(1)}\right\rangle_{\tau^{(l)}},\left\langle\sigma^{(2)}\right\rangle_{\tau^{(l)}},\left\langle\sigma^{(3)}\right\rangle_{\tau^{(l)}}\right)\in\R^3
\]
denote the Bloch vector of \(A\) after the \(l\)th re-uploading layer. Initially,
\[
\tau^{(0)} = |+\rangle\langle+| \quad \Longleftrightarrow \quad \mathbf{r}^{(0)}=\left(r_1^{(0)},r_2^{(0)},r_3^{(0)}\right)=(1,0,0).
\]
At the \(l\)th re-uploading layer, the \(l\)th single-qubit gate \(U^{(l)}\in\text{SU}(2)\) on \(A\) rotates the Bloch vector \(\mathbf{r}^{(l-1)}\) by \(\mathfrak{R}^{(l)}\in\text{SO}(3)\):
\[
\mathbf{r}^{(l-1)} \mapsto \Tilde{\mathbf{r}}^{(l)} = \left(\Tilde{r}_1^{(l)},\Tilde{r}_2^{(l)},\Tilde{r}_3^{(l)}\right) = \mathfrak{R}^{(l)} \mathbf{r}^{(l-1)}.
\]
The state corresponding to \(\Tilde{\mathbf{r}}^{(l)}\) is
\[
\Tilde{\tau}^{(l)} = \frac{1}{2}\left( \I_2 + \Tilde{r}_1^{(l)}X + \Tilde{r}_2^{(l)}Y + \Tilde{r}_3^{(l)}Z \right).
\]
Also, note that
\[
\rho(t) = \begin{pmatrix}
t^2 & t\sqrt{1-t^2}\\
t\sqrt{1-t^2} & 1-t^2
\end{pmatrix} = \frac{1}{2}\left( \I_2 + 2t\sqrt{1-t^2}X + (2t^2-1)Z \right)
\]
and
\[
\text{CNOT}_{B\rightarrow A} = (\I_2)_A \otimes |0\rangle\langle0|_B + X_A \otimes |1\rangle\langle1|_B.
\]
By taking the tensor product \(\Tilde{\tau}^{(l)} \otimes \rho(t)\), applying \(\text{CNOT}_{B\rightarrow A}\), and tracing out system \(B\), the state \(\Tilde{\tau}^{(l)}\) evolves to
\begin{align}
\label{equation:single_param_tau_l}
\tau^{(l)} & = \left( (\I_2)_A \otimes \bra{0}_B \right) \text{CNOT}_{B\rightarrow A} \left( \Tilde{\tau}^{(l)} \otimes \rho(t) \right) \text{CNOT}_{B\rightarrow A} \left( (\I_2)_A \otimes \ket{0}_B \right)\nonumber\\
& + \left( (\I_2)_A \otimes \bra{1}_B \right) \text{CNOT}_{B\rightarrow A} \left( \Tilde{\tau}^{(l)} \otimes \rho(t) \right) \text{CNOT}_{B\rightarrow A} \left( (\I_2)_A \otimes \ket{1}_B \right)\\
& = K_0 \left( \Tilde{\tau}^{(l)} \otimes \rho(t) \right) K_0^\dagger + K_1 \left( \Tilde{\tau}^{(l)} \otimes \rho(t) \right) K_1^\dagger,\nonumber
\end{align}
where the Kraus operators \(K_0\) and \(K_1\) are
\begin{align*}
K_0 & = \left( (\I_2)_A \otimes \bra{0}_B \right) \text{CNOT}_{B\rightarrow A}\\
& = \left( (\I_2)_A \otimes \bra{0}_B \right) \left( (\I_2)_A \otimes |0\rangle\langle0|_B + X_A \otimes |1\rangle\langle1|_B \right)\\
& = (\I_2)_A \otimes \bra{0}_B
\end{align*}
and
\begin{align*}
K_1 & = \left( (\I_2)_A \otimes \bra{1}_B \right) \text{CNOT}_{B\rightarrow A}\\
& = \left( (\I_2)_A \otimes \bra{1}_B \right) \left( (\I_2)_A \otimes |0\rangle\langle0|_B + X_A \otimes |1\rangle\langle1|_B \right)\\
& = X_A \otimes \bra{1}_B.
\end{align*}
After some calculations, we find
\begin{align}
\label{equation:single_param_K0_sandwich}
K_0 \left( \Tilde{\tau}^{(l)} \otimes \rho(t) \right) K_0^\dagger & = \frac{1}{4} \left( \I_2 + \Tilde{r}_1^{(l)}X + \Tilde{r}_2^{(l)}Y + \Tilde{r}_3^{(l)}Z \right) (1+2t^2-1)\nonumber\\
& = \frac{1}{2} t^2 \left( \I_2 + \Tilde{r}_1^{(l)}X + \Tilde{r}_2^{(l)}Y + \Tilde{r}_3^{(l)}Z \right)
\end{align}
and
\begin{align}
\label{equation:single_param_K1_sandwich}
K_1 \left( \Tilde{\tau}^{(l)} \otimes \rho(t) \right) K_1^\dagger & = \frac{1}{4} \left( \I_2 + \Tilde{r}_1^{(l)}X - \Tilde{r}_2^{(l)}Y - \Tilde{r}_3^{(l)}Z \right) (1-2t^2+1)\nonumber\\
& = \frac{1}{2} (1-t^2) \left( \I_2 + \Tilde{r}_1^{(l)}X - \Tilde{r}_2^{(l)}Y - \Tilde{r}_3^{(l)}Z \right).
\end{align}
Let \(\lambda = 2t^2-1 \in (-1, 1)\). Then from Eqs.~\eqref{equation:single_param_tau_l}, \eqref{equation:single_param_K0_sandwich}, and \eqref{equation:single_param_K1_sandwich},
\[
\tau^{(l)} = \frac{1}{2} \left( \I_2 + \Tilde{r}_1^{(l)}X + \lambda \Tilde{r}_2^{(l)}Y + \lambda \Tilde{r}_3^{(l)}Z \right)
\]
and therefore
\[
\mathbf{r}^{(l)} = \left( \Tilde{r}_1^{(l)}, \lambda\Tilde{r}_2^{(l)}, \lambda\Tilde{r}_3^{(l)} \right) = \tilde{\mathbf{r}}^{(l)} \odot (1, \lambda, \lambda),
\]
where \(\odot\) denotes the element-wise (Hadamard) product. That is, the \(y\) and \(z\) components of \(\Tilde{\mathbf{r}}^{(l)}\) are scaled by a factor of \(\lambda\). In summary, the Bloch vector initially at \(\mathbf{r}^{(0)}=(1,0,0)\) undergoes a sequence of alternating rotations and scalings.

\begin{remark}
In general, the Bloch vector undergoes a sequence of more complex transformations. Specifically, let \(U\in\textnormal{SU}(4)\). Then the evolution
\[
\tau^{(l)} \mapsto \tau^{(l+1)} = \Tr_B\!\left(U\left(\tau^{(l)}\otimes\rho\right)U^\dagger\right)
\]
corresponds to
\[
\mathbf{r}^{(l)} \mapsto \mathbf{r}^{(l+1)} = M\mathbf{r}^{(l)}+\mathbf{d},
\]
where \(M\in\R^{3\times3}\) and \(\mathbf{d}\in\R^3\) are given by
\begin{align*}
M_{ij} & = \frac{1}{2} \Tr\!\left[\left(\sigma^{(i)}\otimes\I_2\right)U\left(\sigma^{(j)}\otimes\rho\right)U^\dagger\right],\\
d_i & = \frac{1}{2} \Tr\!\left[\left(\sigma^{(i)}\otimes\I_2\right)U\left(\I_2\otimes\rho\right)U^\dagger\right].
\end{align*}
One can verify that this transformation reduces to a rotation followed by a scaling when the circuit is restricted to Figure~\ref{fig:reupload_restricted}.
\end{remark}

Each \(U^{(l)}\) can be further restricted to the form
\[
R_z(\theta_l) = \begin{pmatrix}
e^{-i\theta_l/2} & 0\\
0 & e^{i\theta_l/2}
\end{pmatrix},
\]
yielding the corresponding Bloch vector rotation
\begin{equation}
\label{equation:bloch_rotation_matrix}
\mathfrak{R}^{(l)} = \begin{pmatrix}
\cos\theta_l & -\sin\theta_l & 0\\
\sin\theta_l& \cos\theta_l & 0\\
0 & 0 & 1
\end{pmatrix}.
\end{equation}
Finally, we introduce learnable parameters \(\mathbf{w}\in\R^3\) and \(b\in\R\), so that the full set of parameters is given by \(\boldsymbol\theta = (\theta_1, \dots , \theta_L, \mathbf{w}, b)\in\R^{L+4}\). The final output is computed via the mapping
\begin{equation}
\label{equation:final_function}
\mathbf{r}^{(L)} \mapsto f_{\boldsymbol\theta}(\lambda) = \mathbf{w} \cdot \mathbf{r}^{(L)} + b \in \R,
\end{equation}
\hl{where $\mathbf{r}^{(L)}$ depends on both $\boldsymbol\theta$ and $\lambda$.} Note that \(f_{\boldsymbol\theta}(\lambda)\) can be expressed as
\[
\langle W\rangle_{\tau^{(L)}}+b = \Tr\!\left(\tau^{(L)}W\right)+b,
\]
where
\[
W = \sum_{i=1}^3 w_i \sigma^{(i)}
\]
is a Hermitian observable. Thus, \(f_{\boldsymbol\theta}(\lambda)\) can be evaluated by computing the expectation value of \(W\) with respect to the state \(\tau^{(L)}\). The expectation values \(\left\langle\sigma^{(1)}\right\rangle\) and \(\left\langle\sigma^{(2)}\right\rangle\) can be accessed using only the ability to perform computational-basis measurements combined with single-qubit Clifford transformations. Moreover, the bias term \(b\) is introduced explicitly for convenience in subsequent analysis, but it can be absorbed into the observable as \(W^\prime=W+b\cdot\I_2\).

We can show that the function defined in Eq.~\eqref{equation:final_function} is capable of representing an arbitrary polynomial in \(\lambda\), given a sufficient number of re-uploading layers \(L\). Consider the following special cases:

\begin{align}
\label{equation:one_hot_param_f_expression}
\boldsymbol\theta = & (0, \dots , 0, \theta_l, 0, \dots , 0, \mathbf{e}_2, 0)\nonumber\\
\rightarrow & \quad \mathbf{r}^{(L)} = \left(\cos\theta_l, \lambda^{L+1-l}\sin\theta_l, 0\right), && f_{\boldsymbol\theta}(\lambda) = \lambda^{L+1-l}\sin\theta_l,\nonumber\\
\boldsymbol\theta = & (0, \dots , 0, \mathbf{e}_2, b)\nonumber\\
\rightarrow & \quad \mathbf{r}^{(L)} = (1,0,0), && f_{\boldsymbol\theta}(\lambda) = b,
\end{align}
where \(\mathbf{e}_2=(0,1,0)\). Since \(f_{\boldsymbol\theta}(\lambda)\) is a polynomial of degree at most \(L\), it can be represented by a coefficient vector \(\mathbf{v}_{\boldsymbol\theta}\in\R^{L+1}\) such that
\[
f_{\boldsymbol\theta}(\lambda) = \sum_{i=1}^{L+1} v_{\boldsymbol\theta,i}\lambda^{i-1}.
\]
Meanwhile, the Jacobian of \(\mathbf{v}_{\boldsymbol\theta}\) at \(\boldsymbol\theta = \boldsymbol\theta_0 = (0, \dots , 0, \mathbf{e}_2, 0)\) is
\begin{equation}
\label{equation:jacobian_binary}
\mathbf{J}_{\mathbf{v}_{\boldsymbol\theta}}(\boldsymbol\theta_0) =
\left[
\begin{NiceArray}{cccccccc}
0 & 0 & \cdots & 0 & 1 & 0 & 0 & 1 \\
0 & 0 & \cdots & 1 & 0 & 0 & 0 & 0 \\
\vdots & \vdots & \ddots & \vdots & \vdots & \vdots & \vdots & \vdots \\
0 & 1 & \cdots & 0 & 0 & 0 & 0 & 0 \\
1 & 0 & \cdots & 0 & 0 & 0 & 0 & 0
\CodeAfter
    \tikz{
        \node[draw=black, thick, fit=(1-5)(5-7), inner sep=2pt] {};
    }
\end{NiceArray}
\right]\in\R^{(L+1)\times(L+4)},
\end{equation}
where the boxed submatrix represents the Jacobian with respect to \(\mathbf{w}\). It is evident that \(\mathbf{J}_{\mathbf{v}_{\boldsymbol\theta}}(\boldsymbol\theta_0)\) has full rank. If we fix $\mathbf{w} = \mathbf{e}_2$ and vary only the remaining $L+1$ parameters
$\boldsymbol\theta^\prime = (\theta_1, \dots, \theta_L, b)$, we obtain the reduced map
\[
\mathbf{v}^\prime_{\boldsymbol\theta^\prime} := \mathbf{v}_{(\theta_1,\dots,\theta_L,\,\mathbf{e}_2,\,b)} .
\]
The Jacobian of \(\mathbf{v}^\prime_{\boldsymbol\theta^\prime}\) at $\boldsymbol\theta' = \mathbf{0}$ is the matrix obtained by removing the boxed submatrix from \(\mathbf{J}_{\mathbf{v}_{\boldsymbol\theta}}(\boldsymbol\theta_0)\) in Eq.~\eqref{equation:jacobian_binary}, which is invertible.

\begin{theorem}[\cite{hormander2015analysis}, Theorem~1.1.7]
\label{theorem:inverse_function_theorem}
If \( f \) is a continuously differentiable function from an open subset \( \Tilde{S} \subset \R^n \) into \( \R^n \), and the derivative \( f'(p) \) is invertible at a point \( p \), then there exist neighborhoods \( S \) of \( p \) in \( \Tilde{S} \) and \( T \) of \( q = f(p) \) such that \( f(S) \subset T \) and \( f : S \to T \) is bijective.
\end{theorem}

Note that \(\mathbf{v}^\prime_{\mathbf{0}}=\mathbf{0}\). Then by Theorem~\ref{theorem:inverse_function_theorem}, for arbitrary \(\mathbf{v}\in\R^{L+1}\), there exists a real constant \(\varepsilon\ne0\) and a point \(\boldsymbol\theta^\prime\in\R^{L+1}\) near \(\mathbf{0}\) such that
\[
\mathbf{v}^\prime_{\boldsymbol\theta^\prime} = \varepsilon\mathbf{v}.
\]
However, we can scale \(\mathbf{v}^\prime_{\boldsymbol\theta^\prime}\) by an arbitrary constant, since
\begin{equation}
\label{equation:scale_poly_vec}
f_{(\theta_1, \dots , \theta_L, c\mathbf{w}, cb)}(\lambda) = c f_{(\theta_1, \dots , \theta_L, \mathbf{w}, b)}(\lambda).
\end{equation}
Choosing \(c=\varepsilon^{-1}\), Eq.~\eqref{equation:scale_poly_vec} implies that there exists a point \(\boldsymbol\theta\in\R^{L+4}\) such that \(\mathbf{v}_{\boldsymbol\theta} = \mathbf{v}\).

\subsection{Re-uploading a single-parameter multi-qubit state}

This section generalizes the single-parameter construction in Section~\ref{section:single_parameter_state} from a single-qubit input to a multi-qubit input. Any \(n\)-qubit density matrix can be represented as
\begin{equation}
\label{eq:general-bloch-rho-expansion}
\rho = \frac{1}{d}\left(\I_d+\sum_{\alpha=1}^{d^2-1}\lambda_\alpha W_\alpha\right),
\end{equation}
where \(W_\alpha\) are generalized Pauli matrices (\(n\)-fold tensor products of \(\sigma^{(0)}\), \(\sigma^{(1)}\), \(\sigma^{(2)}\), and \(\sigma^{(3)}\)). Let \(W_\alpha^+\) and \(W_\alpha^-\) denote the projectors onto the positive and negative eigenspaces of \(W_\alpha\). Because every generalized Pauli is a traceless Hermitian unitary, its only eigenvalues \(\pm1\) must occur equally often, so the \(+1\)- and \(-1\)-eigenspaces each span half of \(\mathcal{H}_d\). Therefore, \(W_\alpha^+\) and \(W_\alpha^-\) can be expressed as
\[
W_\alpha^+ = \frac{1}{2}(\I_d+W_\alpha) = \sum_{\beta=1}^{d/2}\ketbraauto{\psi_\beta^{(\alpha)}}
\]
and
\[
W_\alpha^- = \frac{1}{2}(\I_d-W_\alpha) = \sum_{\gamma=1}^{d/2}\ketbraauto{\phi_\gamma^{(\alpha)}},
\]
where \(\left\{\ketauto{\psi_\beta^{(\alpha)}}\right\}_\beta\) and \(\left\{\ketauto{\phi_\gamma^{(\alpha)}}\right\}_\gamma\) are orthonormal families of states.
Define the multi-qubit controlled unitary
\[
\text{CU}_{B\rightarrow A}^{(\alpha)} = (\I_2)_A\otimes(W_\alpha^+)_B+X_A\otimes(W_\alpha^-)_B
\]
and the Kraus operators
\begin{align*}
K_\beta^{(\alpha+)} & = \left((\I_2)_A\otimes\braauto{\psi_\beta^{(\alpha)}}_B\right) \text{CU}_{B\rightarrow A}^{(\alpha)} = (\I_2)_A\otimes\braauto{\psi_\beta^{(\alpha)}}_B,\\
K_\gamma^{(\alpha-)} & = \left((\I_2)_A\otimes\braauto{\phi_\gamma^{(\alpha)}}_B\right) \text{CU}_{B\rightarrow A}^{(\alpha)} = X_A\otimes\braauto{\phi_\gamma^{(\alpha)}}_B.
\end{align*}
Then the state
\[
\Tilde{\tau}^{(l)}\otimes\rho = \frac{1}{2d}\left( \I_2 + \Tilde{r}_1^{(l)}X + \Tilde{r}_2^{(l)}Y + \Tilde{r}_3^{(l)}Z \right)\otimes\left(\I_d+\sum_{\alpha^\prime=1}^{d^2-1}\lambda_{\alpha^\prime} W_{\alpha^\prime}\right)
\]
evolves to
\begin{equation}
\label{equation:evolution_formula}
\tau^{(l)} = \frac{1}{2} \left( \I_2 + \Tilde{r}_1^{(l)}X  + [\lambda_\alpha\Tilde{r}_2^{(l)}]Y + [\lambda_\alpha\Tilde{r}_3^{(l)}]Z \right),
\end{equation}
where the subsystem labels \(A\) and \(B\) have been omitted for notational simplicity (see Appendix~\ref{section:evolution_formula_derivation} for the derivation). That is, the \(y\) and \(z\) components of \(\Tilde{\mathbf{r}}^{(l)}\) are scaled by a factor of \(\lambda_\alpha\).
If the target function of \(\rho\) depends only on a single parameter \(\lambda_{\alpha^\prime}\), then the proof of universality can be completed by employing the same approach outlined in Section~\ref{section:single_parameter_state} with a target polynomial
\[
f(\lambda_{\alpha^\prime}) = c_0 + \sum_{\omega=1}^\Omega c_\omega \lambda_{\alpha^\prime}^{\omega}.
\]

\hl{
However, this argument is intrinsically single-parameter.  For a fixed choice of
$\alpha$, the channel induced by $\mathrm{CU}_{B\rightarrow A}^{(\alpha)}$
uses the input state $\rho$ only through the single expectation value
$
\lambda_\alpha=\Tr(\rho W_\alpha).
$
Consequently, the resulting one-qubit dynamics only produces expressions that
can be organized as univariate polynomials in $\lambda_\alpha$. It
does not by itself establish universality for a general function
$
f(\lambda_1,\ldots,\lambda_{d^2-1}),
$
because such a function may require arbitrary multivariate monomials, such as
$\lambda_{\alpha_1}\lambda_{\alpha_2}\cdots\lambda_{\alpha_k}$, involving
different generalized Pauli components.  The above construction fails to show how different components can be accessed, stored, mixed, and
multiplied within the same re-uploading architecture to
generate a dense class of multivariate polynomials.
Moreover, it also does not yield a uniform guarantee that the
final readout observable can be chosen with controlled operator norm.
The next section resolves these issues by proving that our architecture can simulate ordinary \emph{classical} data re-uploading.
}

\begin{highlightblock}
\subsection{Re-uploading an arbitrary multi-qubit state}

In this section, we use the re-uploading architecture of Figure~\ref{fig:reupload_general} to simulate an ordinary single-qubit classical data re-uploading circuit whose classical input is $\boldsymbol{\lambda}$.

We use the standard normalization
\[
\Tr(W_\alpha W_\beta)=d\delta_{\alpha\beta},
\qquad
\Tr(W_\alpha)=0.
\]
Therefore the coefficients in Eq.~\eqref{eq:general-bloch-rho-expansion} are recovered as expectation values:
\begin{align}
\Tr(\rho W_\beta)
&=\frac{1}{d}\Tr\!\left[\left(\I_d+\sum_{\alpha=1}^{d^2-1}\lambda_\alpha W_\alpha\right)W_\beta\right] \nonumber\\
&=\frac{1}{d}\left(\Tr(W_\beta)+\sum_{\alpha=1}^{d^2-1}\lambda_\alpha\Tr(W_\alpha W_\beta)\right) \nonumber\\
&=\lambda_\beta .
\label{equation:bloch_coordinate_expectation_rescue}
\end{align}
Since \(W_\alpha\) has eigenvalues in \(\{+1,-1\}\), Eq.~\eqref{equation:bloch_coordinate_expectation_rescue} also implies \(|\lambda_\alpha|\le1\) for every physical input state \(\rho\).
Since each \(W_\alpha\) is Hermitian and unitary, it is involutory:
\[
W_\alpha^\dagger=W_\alpha,
\qquad
W_\alpha^2=\I_d.
\]
Similarly, for each signal Pauli \(\sigma^{(i)}\), \(i\in\{1,2,3\}\),
\[
(\sigma^{(i)})^\dagger=\sigma^{(i)},
\qquad
(\sigma^{(i)})^2=\I_2.
\]

Fix a generalized Bloch coordinate \(\lambda_\alpha\) and a signal Pauli direction \(i\in\{1,2,3\}\).  For a real number \(\delta\), define the \((n+1)\)-qubit unitary
\begin{equation}
\label{equation:weak_collision_unitary_rescue}
U_{\alpha,i}(\delta)
=
\exp\!\left(-i\delta\,\sigma_A^{(i)}\otimes (W_\alpha)_B\right).
\end{equation}
This acts only on the signal register \(A\) and the current copy of $\rho$ in register \(B\).  If
\[
H_{\alpha,i}=\sigma_A^{(i)}\otimes(W_\alpha)_B,
\]
then
\[
\Tr(H_{\alpha,i})=\Tr(\sigma^{(i)})\Tr(W_\alpha)=0,
\]
so \(\det(U_{\alpha,i}(\delta))=\exp(-i\delta\Tr(H_{\alpha,i}))=1\).  Thus \(U_{\alpha,i}(\delta)\in\mathrm{SU}(2d)\).  Also,
\[
H_{\alpha,i}^2=(\sigma_A^{(i)})^2\otimes ((W_\alpha)_B)^2=(\I_2)_A\otimes(\I_d)_B.
\]
Expanding the exponential in Eq.~\eqref{equation:weak_collision_unitary_rescue} gives
\begin{equation}
U_{\alpha,i}(\delta)
=
\cos\delta\, (\I_2)_A\otimes(\I_d)_B
-i\sin\delta\,\sigma_A^{(i)}\otimes(W_\alpha)_B
\label{equation:weak_collision_cos_sin_rescue}
\end{equation}
(see Appendix~\ref{sec:exp-expand}).

Let \(\tau\in\mathcal{D}(\mathcal{H}_2)\) be the current state of the signal register.  The re-uploading layer induced by Eq.~\eqref{equation:weak_collision_unitary_rescue} is the channel
\begin{equation}
\label{equation:weak_collision_channel_definition_rescue}
\Phi_{\alpha,i}^{(\delta)}(\tau)
=
\Tr_B\!\left[
U_{\alpha,i}(\delta)(\tau\otimes\rho)U_{\alpha,i}(\delta)^\dagger
\right].
\end{equation}
For compactness, set
\[
c_\delta=\cos\delta,
\qquad
s_\delta=\sin\delta,
\qquad
S_i=\sigma^{(i)}.
\]
Using Eq.~\eqref{equation:weak_collision_cos_sin_rescue} and suppressing subsystem labels, we have
\begin{align}
U_{\alpha,i}(\delta)(\tau\otimes\rho)U_{\alpha,i}(\delta)^\dagger
&=\left(c_\delta\I-i s_\delta S_i\otimes W_\alpha\right)
(\tau\otimes\rho)
\left(c_\delta\I+i s_\delta S_i\otimes W_\alpha\right) \nonumber\\
&=c_\delta^2(\tau\otimes\rho)
+s_\delta^2(S_i\tau S_i\otimes W_\alpha\rho W_\alpha) \nonumber\\
&\quad
+i c_\delta s_\delta
\left[(\tau S_i\otimes\rho W_\alpha)-(S_i\tau\otimes W_\alpha\rho)\right].
\label{equation:weak_collision_expanded_joint_state_rescue}
\end{align}
Now take the partial trace over register \(B\).  The first term gives
\[
\Tr_B(\tau\otimes\rho)=\tau\Tr(\rho)=\tau.
\]
The second term gives
\begin{align*}
\Tr_B(S_i\tau S_i\otimes W_\alpha\rho W_\alpha)
&=S_i\tau S_i\Tr(W_\alpha\rho W_\alpha)\\
&=S_i\tau S_i\Tr(\rho W_\alpha^2)\\
&=S_i\tau S_i.
\end{align*}
For the two cross terms, Eq.~\eqref{equation:bloch_coordinate_expectation_rescue} gives
\begin{align*}
\Tr_B(\tau S_i\otimes\rho W_\alpha)
&=\tau S_i\Tr(\rho W_\alpha)=\lambda_\alpha\tau S_i,\\
\Tr_B(S_i\tau\otimes W_\alpha\rho)
&=S_i\tau\Tr(W_\alpha\rho)=\lambda_\alpha S_i\tau.
\end{align*}
Substituting these identities into Eq.~\eqref{equation:weak_collision_expanded_joint_state_rescue}, we obtain
\begin{align}
\Phi_{\alpha,i}^{(\delta)}(\tau)
&=c_\delta^2\tau+s_\delta^2S_i\tau S_i
+i c_\delta s_\delta\lambda_\alpha(\tau S_i-S_i\tau) \nonumber\\
&=c_\delta^2\tau+s_\delta^2S_i\tau S_i
-i c_\delta s_\delta\lambda_\alpha[S_i,\tau].
\label{equation:exact_weak_collision_channel_rescue}
\end{align}
This identity shows that a single re-uploading layer produces, to first order in \(\delta\), a unitary rotation of the signal qubit whose angle is proportional to \(\lambda_\alpha\).

Indeed, Eq.~\eqref{equation:exact_weak_collision_channel_rescue} gives
\begin{align}
\Phi_{\alpha,i}^{(\delta)}(\tau)
&=\tau-i\delta\lambda_\alpha[S_i,\tau]
+\delta^2(S_i\tau S_i-\tau)
+O(\delta^3).
\label{equation:weak_collision_first_order_rescue}
\end{align}
The term \(-i\delta\lambda_\alpha[S_i,\tau]\) is exactly the first-order action of the data-dependent rotation
\begin{equation}
\label{equation:ideal_classical_rotation_gate_rescue}
R_{\alpha,i}(t;\boldsymbol\lambda)
=
\exp(-it\lambda_\alpha S_i).
\end{equation}
Let
\begin{equation}
\label{equation:ideal_classical_rotation_channel_rescue}
\mathcal{R}_{\alpha,i}^{(t)}(\tau)
=
R_{\alpha,i}(t;\boldsymbol\lambda)\tau R_{\alpha,i}(t;\boldsymbol\lambda)^\dagger.
\end{equation}
Since \(S_i^2=\I_2\),
\[
R_{\alpha,i}(t;\boldsymbol\lambda)
=
\cos(t\lambda_\alpha)\I_2-i\sin(t\lambda_\alpha)S_i.
\]
Hence the classical data-encoding rotation is
\begin{align}
\mathcal{R}_{\alpha,i}^{(t)}(\tau)
&=\cos^2(t\lambda_\alpha)\tau
+\sin^2(t\lambda_\alpha)S_i\tau S_i
-i\sin(t\lambda_\alpha)\cos(t\lambda_\alpha)[S_i,\tau].
\label{equation:exact_ideal_rotation_channel_rescue}
\end{align}
Comparing Eq.~\eqref{equation:exact_weak_collision_channel_rescue} at \(\delta=t\) with Eq.~\eqref{equation:exact_ideal_rotation_channel_rescue}, the first-order terms agree.  The difference begins at second order in the small interaction angle.  More explicitly, for \(|\lambda_\alpha|\le1\),
\begin{align}
\Phi_{\alpha,i}^{(\delta)}(\tau)-\mathcal{R}_{\alpha,i}^{(\delta)}(\tau)
&=\left(\cos^2\delta-\cos^2(\delta\lambda_\alpha)\right)\tau + \left(\sin^2\delta-\sin^2(\delta\lambda_\alpha)\right)S_i\tau S_i \nonumber\\
&\quad-i\left(\lambda_\alpha\sin\delta\cos\delta-
\sin(\delta\lambda_\alpha)\cos(\delta\lambda_\alpha)\right)[S_i,\tau],
\label{equation:weak_minus_ideal_rescue}
\end{align}
and this difference is uniformly \(O(\delta^2)\) over all input states \(\rho\) and all signal states \(\tau\): there exists a constant \(C>0\), independent of \(\rho\), \(\tau\), \(\alpha\), and \(i\), such that for all \(|\delta|\le1\),
\begin{equation}
\label{equation:single_step_error_bound_rescue}
\norm{\Phi_{\alpha,i}^{(\delta)}(\tau)-\mathcal{R}_{\alpha,i}^{(\delta)}(\tau)}_1
\le C\delta^2.
\end{equation}

A finite-angle classical data-dependent rotation is obtained from many re-uploading layers.  Fix a desired rotation angle parameter \(t\in\R\), choose an integer \(N\ge1\), and set \(\delta=t/N\).  The re-uploading architecture can apply the channel \(\Phi_{\alpha,i}^{(t/N)}\) repeatedly \(N\) times, each time using a fresh copy of \(\rho\) in register \(B\).  Since the ideal channels satisfy
\[
\left(\mathcal{R}_{\alpha,i}^{(t/N)}\right)^N
=
\mathcal{R}_{\alpha,i}^{(t)},
\]
we have
\begin{equation*}
\lim_{N\rightarrow\infty}
\left(\Phi_{\alpha,i}^{(t/N)}\right)^N(\tau)
=
\mathcal{R}_{\alpha,i}^{(t)}(\tau)
\end{equation*}
uniformly in \(\rho\) and \(\tau\). Therefore, the quantum data re-uploading architecture can approximate the classical data-dependent single-qubit rotation
\[
\tau\mapsto
\exp(-it\lambda_\alpha S_i)\tau\exp(it\lambda_\alpha S_i)
\]
by using many weak interactions.

We next embed this primitive into an ordinary single-qubit classical data re-uploading circuit.  Consider a classical data re-uploading model whose classical input is
\[
\boldsymbol\lambda=(\lambda_1,\ldots,\lambda_{d^2-1})\in\R^{d^2-1}.
\]
An instance of such a model with finite depth has the form
\begin{equation}
\label{equation:classical_reuploading_circuit_rescue}
\mathcal{C}_{\boldsymbol\theta}(\boldsymbol\lambda)
=
\mathcal{V}^{(M)}\circ\mathcal{R}_{\alpha_M,i_M}^{(t_M)}\circ
\mathcal{V}^{(M-1)}\circ\cdots\circ
\mathcal{R}_{\alpha_1,i_1}^{(t_1)}\circ\mathcal{V}^{(0)},
\end{equation}
where each \(\mathcal{V}^{(m)}\) is an input-independent single-qubit unitary channel,
\[
\mathcal{V}^{(m)}(\tau)=V^{(m)}\tau(V^{(m)})^\dagger,
\qquad
V^{(m)}\in\mathrm{SU}(2),
\]
and each \(\mathcal{R}_{\alpha_m,i_m}^{(t_m)}\) is defined as in Eq.~\eqref{equation:ideal_classical_rotation_channel_rescue}.  The symbol \(\boldsymbol\theta\) denotes the continuous trainable parameters appearing in the circuit and readout.  The Pauli directions \(i_m\) and the coordinates \(\alpha_m\) are discrete architecture choices. This proves that every finite ordinary classical data re-uploading circuit on the generalized Bloch vector \(\boldsymbol\lambda\) can be approximated by a single quantum data re-uploading circuit. Therefore, the quantum data re-uploading architecture inherits the universal approximation capability of ordinary single-qubit classical data re-uploading on the compact generalized Bloch-vector domain \cite{perez2025universal}.

\begin{theorem}[Universality of quantum data re-uploading]
\label{theorem:quantum_data_reuploading_universality}
Let \(n\in\mathbb{N}\), let \(d=2^n\), and set \(m=d^2-1\).  For
\(\rho\in\mathcal{D}(\mathcal{H}_d)\), write
\[
    \boldsymbol\lambda(\rho)
    =\bigl(\lambda_1(\rho),\ldots,\lambda_m(\rho)\bigr),
    \qquad
    \lambda_\alpha(\rho)=\Tr(\rho W_\alpha),
\]
and let
\[
    \Lambda_d
    =\{\boldsymbol\lambda(\rho):\rho\in\mathcal{D}(\mathcal{H}_d)\}
    \subset[-1,1]^m .
\]
Consider the following input access and circuit model.

\begin{enumerate}
    \item The input is given by fresh copy access to an unknown state
    \(\rho\in\mathcal{D}(\mathcal{H}_d)\).  Equivalently, before each upload the input register
    \(B\simeq\mathcal{H}_d\) can be reset to \(\rho\).  No classical
    description of \(\rho\) or of \(\boldsymbol\lambda(\rho)\) is assumed.

    \item The signal register is a single qubit
    \(A\simeq\mathcal{H}_2\), initialized in \(\tau^{(0)}=|0\rangle\langle0|\).
    The allowed gates are arbitrary single-qubit
    unitaries \(V\in\mathrm{SU}(2)\) on \(A\) and the upload gates
    \[
        U_{\alpha,i}(\delta)
        =\exp\!\left(-i\delta\,\sigma_A^{(i)}\otimes(W_\alpha)_B\right),
        \qquad
        \alpha\in\{1,\ldots,m\},\quad i\in\{1,2,3\},
        \quad \delta\in\R .
    \]

    \item After each upload gate, register \(B\) is discarded or reset,
    while \(A\) is kept coherent.

    \item Only the final signal qubit \(A\) is measured.  The allowed
    readout is an affine combination of single-qubit Pauli expectations,
    \[
        h(\rho)
        =b_{\rm out}+\sum_{i=1}^3 a_i
        \Tr\!\left(\sigma^{(i)}\tau_A^{\rm out}(\rho)\right)
        =b_{\rm out}+\Tr\!\left(O_{\rm out}\tau_A^{\rm out}(\rho)\right),
    \]
    where
    $
        O_{\rm out}=\sum_{i=1}^3 a_i\sigma^{(i)}\in\operatorname{Herm}(\mathcal{H}_2).
    $
\end{enumerate}

Then the model is a universal approximator for continuous functions of
\(\rho\) in the uniform norm.  More precisely, for every bounded continuous
function \(F:\mathcal{D}(\mathcal{H}_d)\to\R\) and every \(\varepsilon>0\),
there exist a finite number \(L_Q\) of uploads, choices of
\((\alpha,i,\delta)\) for those uploads, single-qubit
unitaries on \(A\), and a readout \((O_{\rm out},b_{\rm out})\) such that
\[
    \sup_{\rho\in\mathcal{D}(\mathcal{H}_d)}
    \left|
        b_{\rm out}+\Tr\!\left(O_{\rm out}\tau_A^{\rm out}(\rho)\right)
        -F(\rho)
    \right|<\varepsilon
\]
and
\begin{equation}
\label{eq:readout-obs-upper-bound}
    \norm{O_{\rm out}}_\infty
    \le
    \frac{1}{2}
    \left(
        \sup_{\rho\in\mathcal{D}(\mathcal{H}_d)} F(\rho)
        -
        \inf_{\rho\in\mathcal{D}(\mathcal{H}_d)} F(\rho)
    \right).
\end{equation}
Furthermore, the approximation can be chosen by first selecting a finite single-qubit classical re-uploading circuit of the form
given in Eq.~\eqref{equation:classical_reuploading_circuit_rescue}, and then replacing each data-dependent rotation
\(\mathcal{R}_{\alpha_j,i_j}^{(t_j)}\) by \(N_j\) weak upload layers
\((\Phi_{\alpha_j,i_j}^{(t_j/N_j)})^{N_j}\).  If \(N_j\ge |t_j|\) for all \(j\), the simulated
quantum channel
\[
\widehat{\mathcal{C}}_{\boldsymbol\theta,\mathbf{N}}^{\rho}
=
\mathcal{V}^{(M)}\circ(\Phi_{\alpha_M,i_M}^{(t_M/N_M)})^{N_M}\circ
\mathcal{V}^{(M-1)}\circ\cdots\circ
(\Phi_{\alpha_1,i_1}^{(t_1/N_1)})^{N_1}\circ\mathcal{V}^{(0)}
\]
satisfies the uniform trace-norm simulation bound
\begin{equation}
\label{equation:formal_simulation_bound}
    \sup_{\rho\in\mathcal{D}(\mathcal{H}_d)}
    \sup_{\tau\in\mathcal{D}(\mathcal{H}_2)}
    \left\|
        \widehat{\mathcal{C}}_{\boldsymbol\theta,\mathbf{N}}^{\rho}(\tau)
        -\mathcal{C}_{\boldsymbol\theta}(\boldsymbol\lambda(\rho))(\tau)
    \right\|_1
    \le
    6\sum_{j=1}^M \frac{t_j^2}{N_j} .
\end{equation}
Consequently, the total number of copies used is
\(L_Q=\sum_{j=1}^M N_j\).
\end{theorem}

\begin{proof}
See Appendix~\ref{section:formal_universality}.
\end{proof}

\begin{remark}
\label{rem:existential}
The result above should be interpreted as an existential
width-efficient expressivity statement, not as an efficient $\mathrm{SU}(2d)$ circuit-synthesis theorem.
In other words, we do not prove that the number of upload layers, elementary gates, or sample complexity scale efficiently for arbitrary tasks.
Thus the theorem establishes universality with small qubit width,
whereas the design of explicit and depth-efficient circuit families
for useful classes of functions remains a separate problem.
\end{remark}
\end{highlightblock}

\section{Interpretation as an entangling measurement}
\label{sec:explicit_model_interpretation}

\begin{figure}
    \centering
    \begin{subfigure}[b]{0.9\linewidth}
        \centering
        \includegraphics[width=\linewidth]{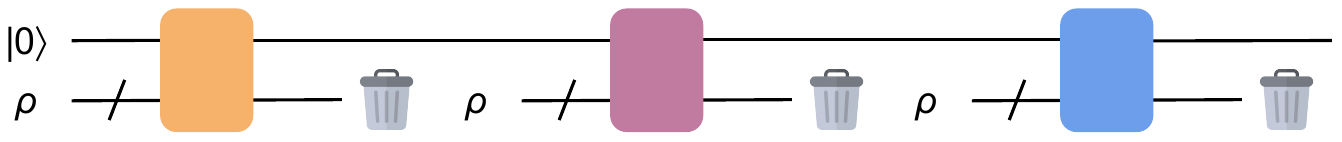}
        \caption{\centering}
        \vspace{1em}
        \label{fig:explicit_model_a}
    \end{subfigure}
    \begin{subfigure}[b]{0.54\linewidth}
        \centering
        \includegraphics[width=\linewidth]{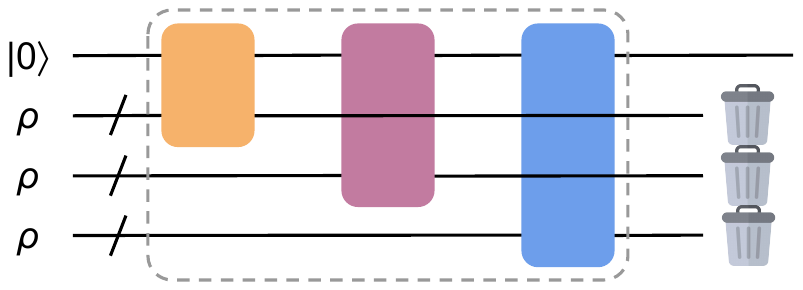}
        \caption{\centering}
        \label{fig:explicit_model_b}
    \end{subfigure}
    \caption{Two equivalent representations of our re-uploading model.}
    \label{fig:explicit_model}
\end{figure}

By the deferred measurement principle, the circuit in Figure~\ref{fig:reupload_general} with \(L\) re-uploading layers is equivalent to an \((Ln+1)\)-qubit circuit that prepares \(L\) copies of \(\rho\), applies a sequence of \(L\) \((n+1)\)-qubit unitaries, and performs measurements on the signal register (see Figure~\ref{fig:explicit_model}).

This reinterpretation provides a perspective closely related to Ref.~\cite{huang2022quantum}. In both cases, quantum memory enables entangling measurements across multiple copies of a state and provides capabilities far beyond sequential single-copy measurements with classical post-processing.
\hl{Since the signal register serves as the quantum memory that carries information between successive uploads, we expect the potential advantage of the architecture to become more pronounced when this register is enlarged from a single qubit to a larger system.}
In our architecture (Figure~\ref{fig:explicit_model_b}), the unitary transformation enclosed in the dashed box can be absorbed into the measurement. This enables the approximation of arbitrary continuous functions of the target quantum input state (given sufficient copies and depth), a task that is more difficult when restricted to local measurements followed by classical post-processing.

To efficiently train the circuit in Figure~\ref{fig:explicit_model_b}, one can apply parameter-shift rules, which express quantum derivatives via function evaluations at shifted inputs \cite{mitarai2018quantum, schuld2019evaluating}. Given the logical equivalence to the re-uploading model in Figure~\ref{fig:explicit_model_a}, and our focus on observables in system \(A\), these rules can be applied just as effectively within the re-uploading formulation.
However, in our numerical simulations (Section~\ref{sec:experiments}), parameter-shift rules do not directly apply due to the general full-unitary parameterizations. We instead employ automatic differentiation to evaluate the gradients.

We briefly discuss the case of machine learning on a sequence of quantum states, i.e., on approximating arbitrary functions of
\[
\mathcal{S} = \left( \rho^{(1)}, \dots, \rho^{(N)} \right).
\]
It follows directly from the previous sections that re-uploading the tensor product state
\[
\rho_\text{prod} = \rho^{(1)} \otimes \cdots \otimes \rho^{(N)}
\]
enables approximation of any bounded continuous function of the sequence \(\mathcal{S}\). However, qubit resources can be used more efficiently. Instead of uploading the full tensor product at once, it suffices to upload a single \(\rho^{(i)}\) at each layer while still preserving universality. To see this, let \(\rho^{(i)}\in\mathcal{D}(\mathcal{H}_{d_i})\) with Bloch vector coefficients
\[
\boldsymbol\lambda^{(i)} = \left( \lambda_\alpha^{(i)} \right)_{\alpha=1}^{d_i^2-1}.
\]
The joint state \(\rho_\text{prod}\) is fully determined by the concatenated coefficient vector
\[
\tilde{\boldsymbol\lambda} = \concat_{i=1}^N \boldsymbol\lambda^{(i)} \in \R^{\sum_{i=1}^N d_i^2 - N}.
\]
Consequently, any bounded continuous function of \(\rho_\text{prod}\) can be uniformly approximated by a polynomial in the variables of \(\tilde{\boldsymbol\lambda}\).
\hl{At each layer, we may select a variable $\lambda_\alpha^{(i)}$ and approximate the corresponding data-dependent rotation of the signal qubit.}
By iterating across layers and across different states \(\rho^{(i)}\), one recovers the same polynomial approximation structure as in the single-state case, requiring at most \(1+\lceil\max_i\log_2d_i\rceil\) qubits.

This perspective also resonates with the process learning setting discussed in Ref.~\cite{huang2022quantum}, where access to quantum memory enables joint measurements across a sequence of states, providing a fundamental advantage over independent single-copy measurements.

\section{Experiments}
\label{sec:experiments}

\begin{figure}
    \centering
    \begin{subfigure}[b]{\linewidth}
        \centering
        \includegraphics[width=0.12\linewidth]{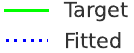}
        \par  
        \vspace{1em}
        \begin{minipage}{\linewidth}
            \centering
            \raisebox{-.5\height}{\includegraphics[width=0.32\linewidth]{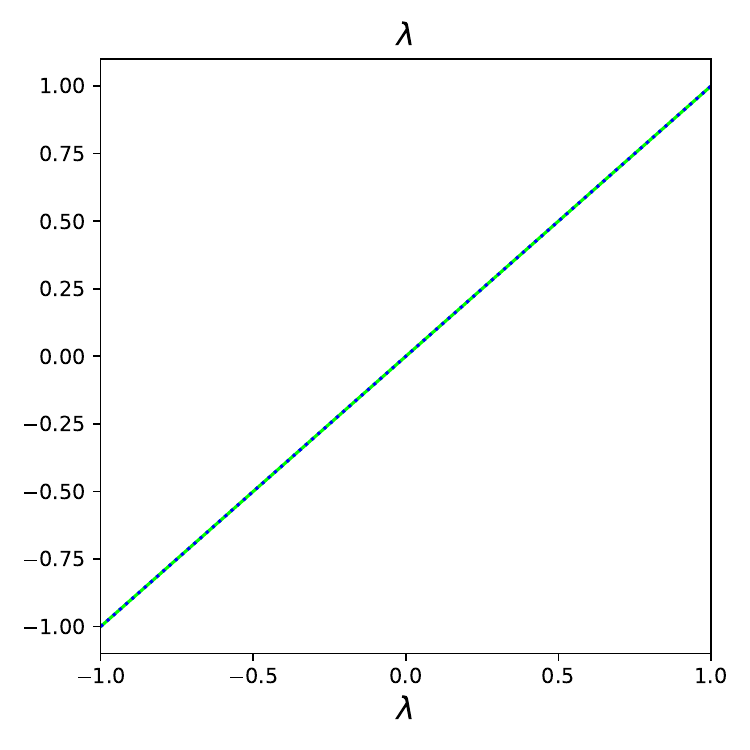}}
            \hfill
            \raisebox{-.5\height}{\includegraphics[width=0.32\linewidth]{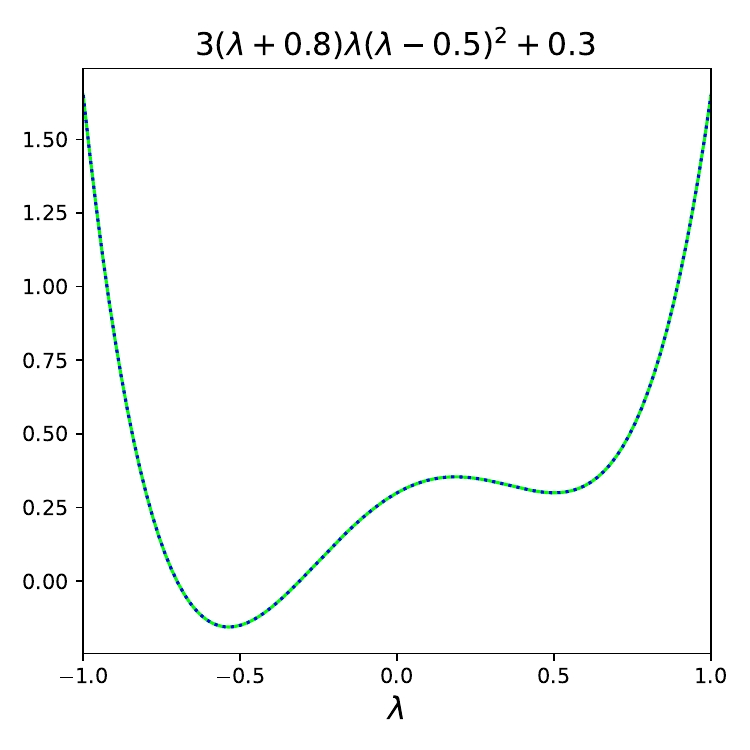}}
            \hfill
            \raisebox{-.5\height}{\includegraphics[width=0.32\linewidth]{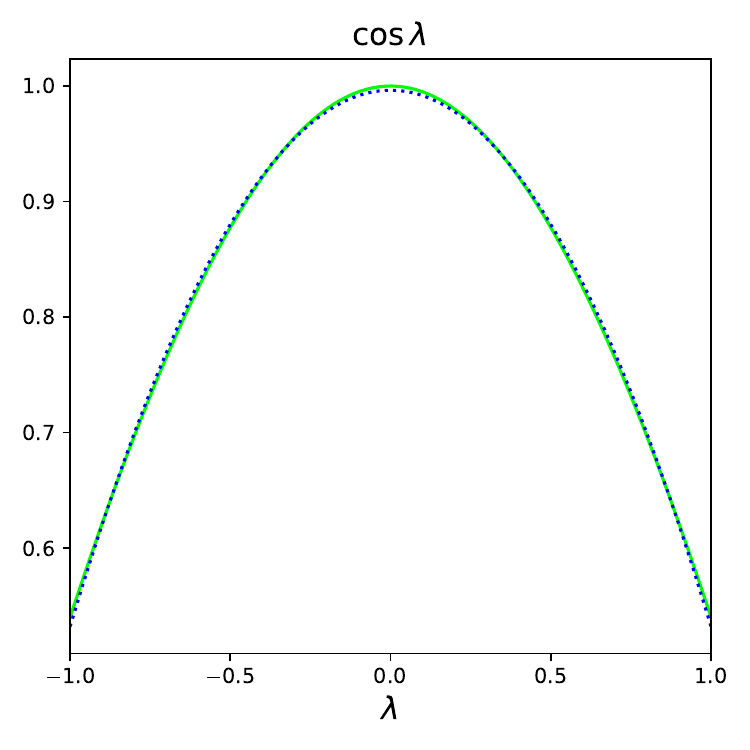}}
        \end{minipage}
        \caption{\centering}
        \vspace{1em}
        \label{fig:fitting_train}
    \end{subfigure}
    \begin{subfigure}[b]{\linewidth}
        \centering
        \begin{minipage}{\linewidth}
            \centering
            \includegraphics[width=0.24\linewidth]{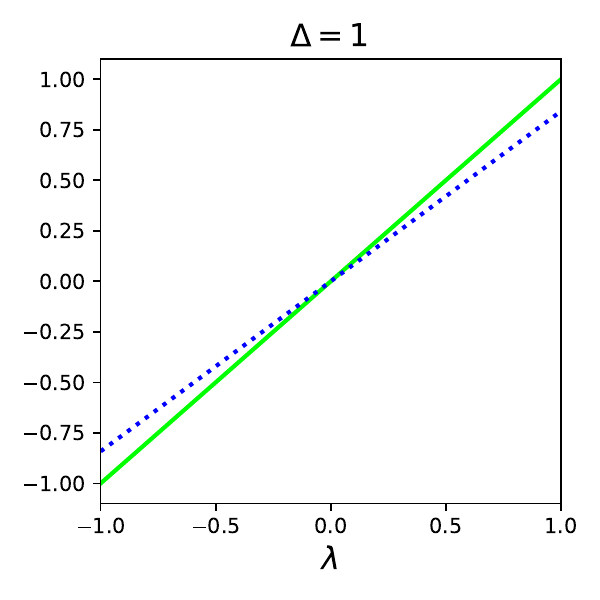}
            \hfill
            \includegraphics[width=0.24\linewidth]{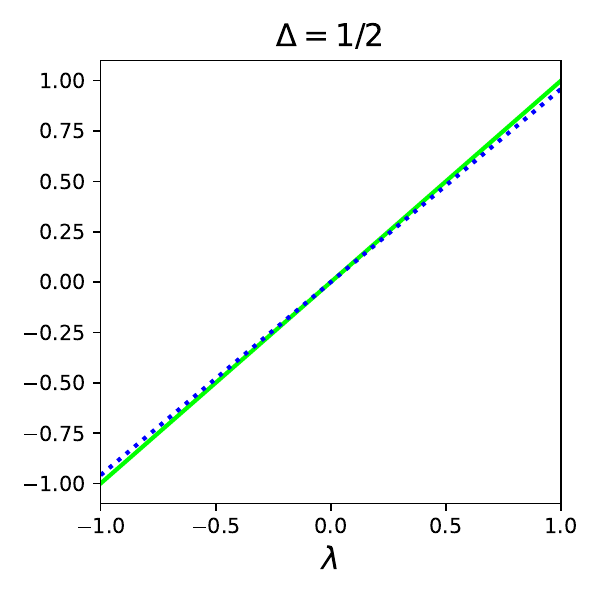}
            \hfill
            \includegraphics[width=0.24\linewidth]{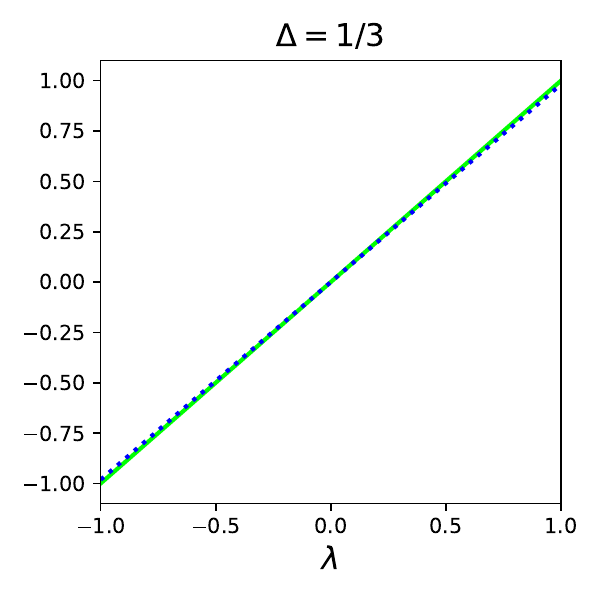}
            \hfill
            \includegraphics[width=0.24\linewidth]{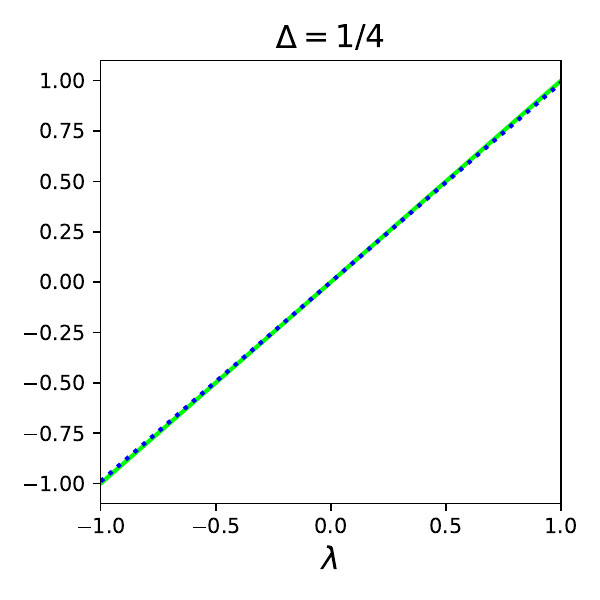}
        \end{minipage}
        \begin{minipage}{\linewidth}
            \centering
            \includegraphics[width=0.24\linewidth]{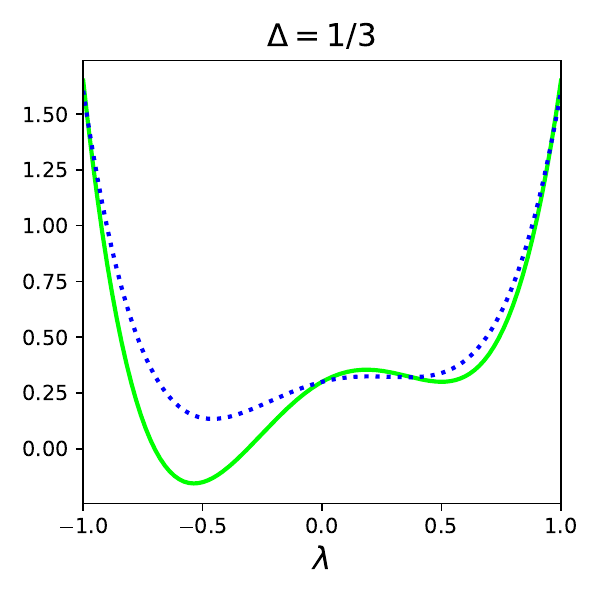}
            \hfill
            \includegraphics[width=0.24\linewidth]{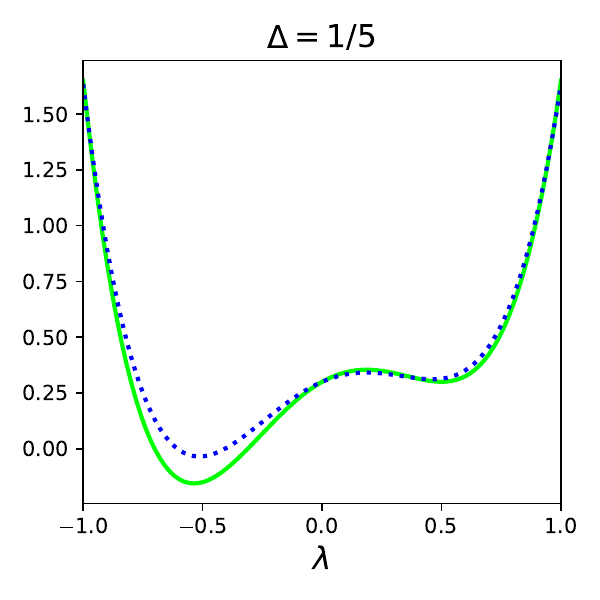}
            \hfill
            \includegraphics[width=0.24\linewidth]{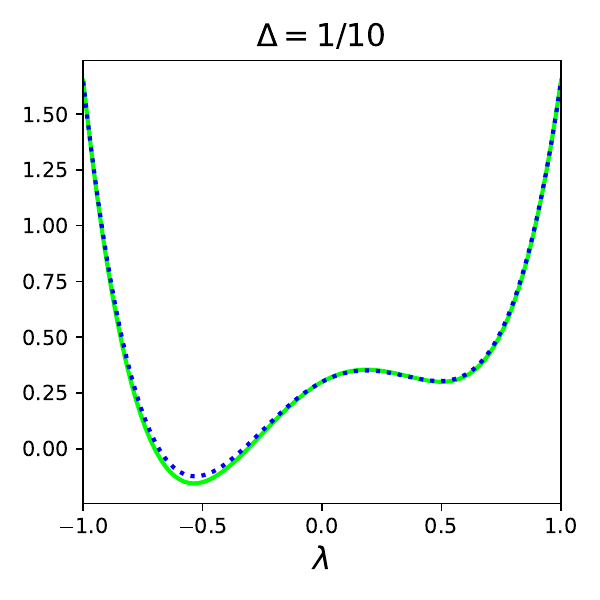}
            \hfill
            \includegraphics[width=0.24\linewidth]{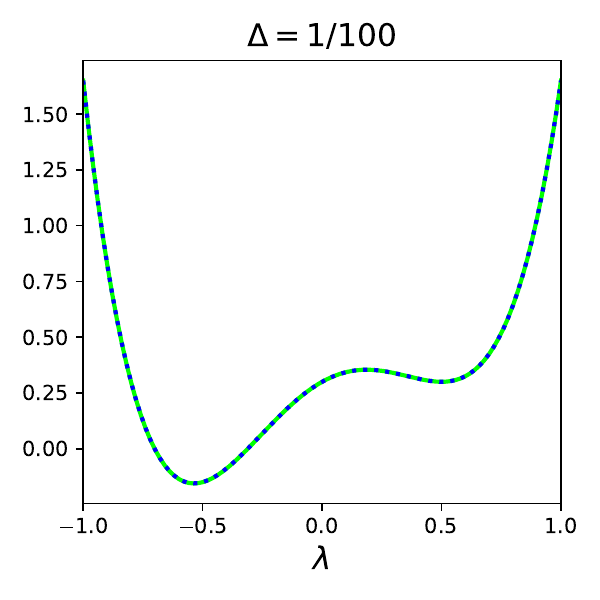}
        \end{minipage}
        \caption{Top row: \(\lambda\). Bottom row: \(3(\lambda+0.8)\lambda(\lambda-0.5)^2+0.3\).}
        \label{fig:fitting_notrain}
    \end{subfigure}
    \caption{Polynomial fitting results for the state in Eq.~\eqref{equation:psi_t_vector}, where \(\lambda=2t^2-1\). Solid green lines denote the target functions, and dotted blue lines denote the fitted results obtained with our re-uploading model. (a) Results for training the restricted re-uploading circuit in Figure~\ref{fig:reupload_restricted}. The target function to be fitted is shown at the top of each plot. From left to right: plots with \(L=1\), \(L=4\), and \(L=2\) layers, respectively. (b) Approximation without training, using Eq.~\eqref{equation:delta_parameters} for different values of \(\Delta\).}
    \label{fig:poly_fitting}
\end{figure}

While the preceding sections establish universality through infinitesimal and existence arguments, such proofs do not imply uniqueness of the corresponding parameter solutions. In practice, training the re-uploading circuit may yield different parameter configurations that realize the same target function. To investigate this trainability aspect beyond pure existence proofs, we conduct a series of experiments in this section, including both regression and classification tasks.

\begin{highlightblock}
\paragraph{Implementation details.}
In all experiments, rather than using a specific circuit ansatz, we employ a full unitary parameterization via
$
U = \exp(iH),
$
where
$
H \in \operatorname{Herm}(\mathcal{H}_{2d}).
$
This full-unitary parameterization is adopted for numerical flexibility rather than hardware efficiency (see Remark~\ref{rem:existential}).
Training was implemented in PyTorch using full-batch optimization. We used the Adam optimizer with learning rate \(10^{-2}\). As a heuristic, all real parameters were initialized independently from \(\mathcal{N}(0,1)\). The signal qubit was initialized to \(\ket{0}\) in all runs.
For polynomial fitting, the regression grid consists of \(201\) evenly spaced points \(\lambda_i\in[-1,1]\). The regression loss is the mean squared error.
The models were trained for \(10^4\) epochs.
For all classification experiments, the training and test set sizes are \(N_{\rm train}=1000\) and \(N_{\rm test}=500\). The model is trained for \(1000\) epochs against binary labels \(y_i\in\{0,1\}\) using the mean squared error.
Code is available in \cite{cha2026reuploadcode}.
\end{highlightblock}

\subsection{Polynomial fitting}

As a simple example, we trained the restricted re-uploading circuit in Figure~\ref{fig:reupload_restricted} using the input states in Eq.~\eqref{equation:psi_t_vector}. The only trainable elements are sequences of single-qubit unitaries, and our targets are three univariate functions in the variable \(\lambda=2t^2-1\) (see Figure~\ref{fig:poly_fitting}). The results are shown in Figure~\ref{fig:fitting_train}.

We provide an alternative perspective on the re-uploading model's approximation capabilities. Let \(\mathbf{v}\in\R^{L+1}\) and define
\begin{equation}
\label{equation:delta_parameters}
\boldsymbol{\theta}(\mathbf{v};\Delta) = \left(v_{L+1}\Delta,\dots,v_2\Delta, \frac{\mathbf{e}_2}{\Delta}, v_1\right) \in \R^{L+4}.
\end{equation}
Then from Eq.~\eqref{equation:one_hot_param_f_expression}, we see that
\[
\lim_{\Delta\rightarrow0} f_{\boldsymbol{\theta}(\mathbf{v};\Delta)} = \sum_{i=1}^{L+1}v_i\lambda^{i-1}.
\]
Note that the parameters in Eq.~\eqref{equation:delta_parameters} are for illustrative purposes only, as the true polynomial coefficients \(\mathbf{v}\) are unknown in practical scenarios. If \(\mathbf{v}\) is known, then the parameters in Eq.~\eqref{equation:delta_parameters} can be directly computed by fixing \(\Delta\), without requiring training. However, taking \(\Delta\rightarrow0\) is impractical, since the factor \(1/\Delta\) diverges and consequently drives the weight vector \(\mathbf{w}\) in Eq.~\eqref{equation:final_function} to blow up. This leads to poor numerical behavior, especially when only a limited number of measurement shots are available to estimate expectation values of observables. In practice, we observe that training the parameters naturally avoids this small-\(\Delta\) regime, while still achieving comparable approximation performance to that obtained from Eq.~\eqref{equation:delta_parameters} with extremely small values of \(\Delta\). Approximation results are shown in Figure~\ref{fig:fitting_notrain}.

\subsection{Purity classification}

\begin{figure}[t]
    \centering
    \begin{subfigure}[b]{0.49\linewidth}
        \includegraphics[width=\linewidth]{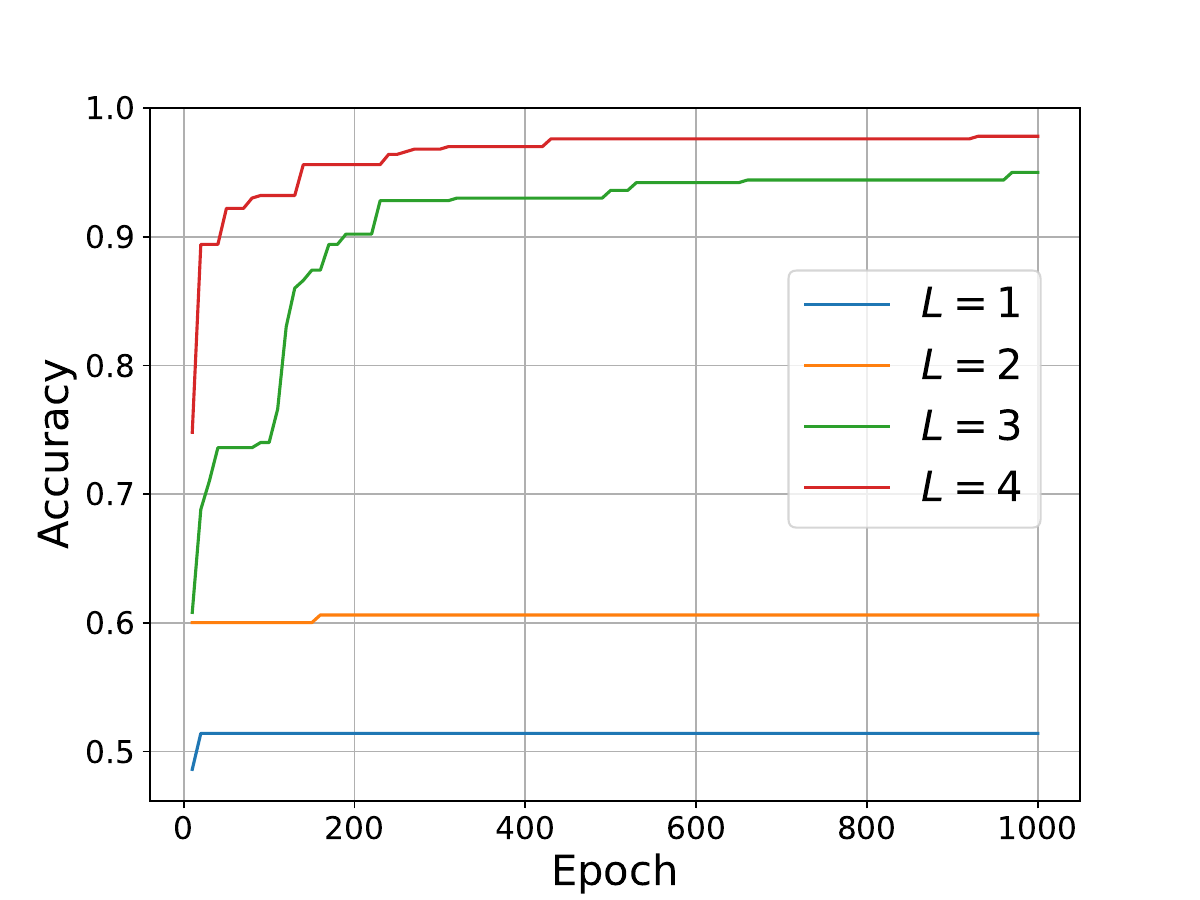}
        \caption{\centering}
        \label{fig:purity_classification_accuracy_curve}
    \end{subfigure}
    \hfill
    \begin{subfigure}[b]{0.49\linewidth}
        \includegraphics[width=\linewidth]{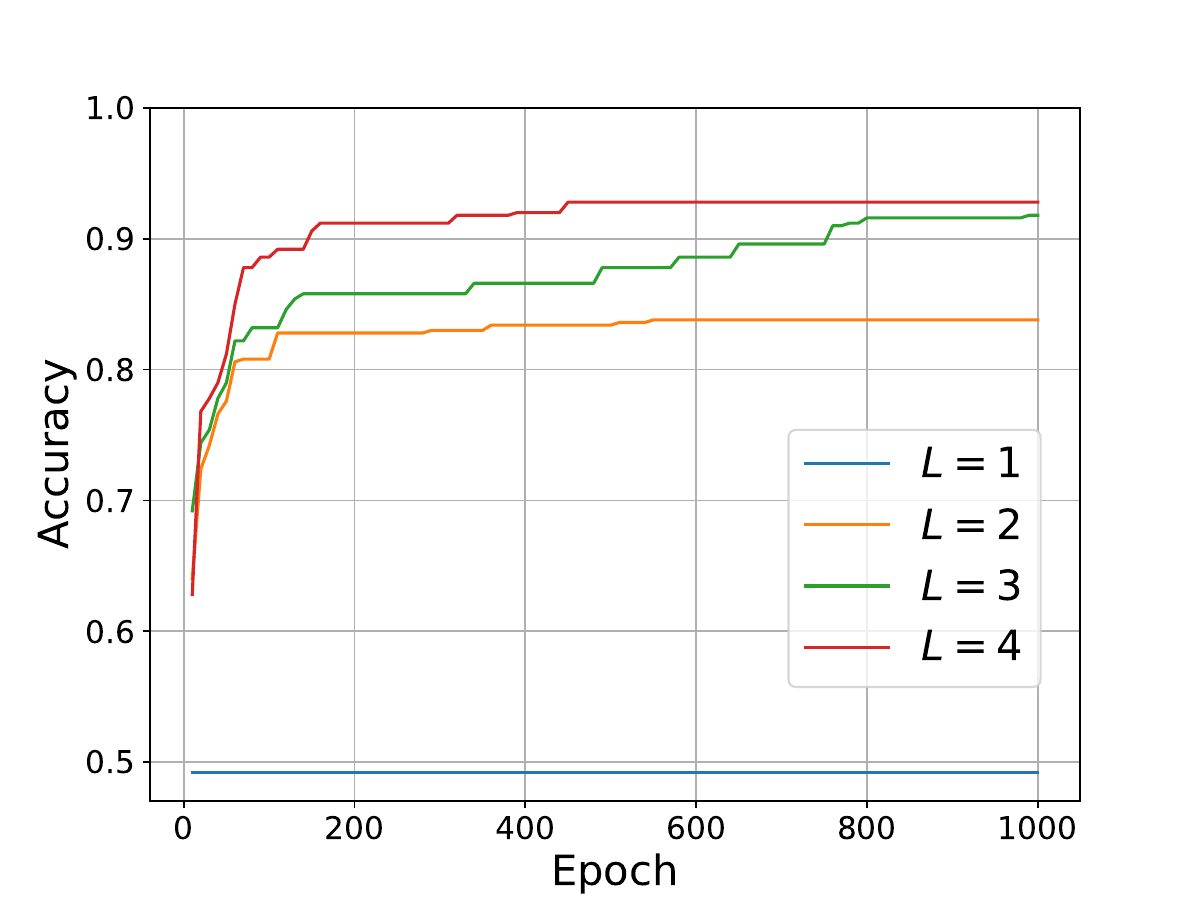}
        \caption{\centering}
        \label{fig:entropy_classification_accuracy_curve}
    \end{subfigure}
    \caption{Test accuracy curves for (a) purity classification and (b) entanglement entropy classification with \(L=1,2,3,4\).}
\end{figure}

\begin{figure}[t]
    \centering
    \includegraphics[width=0.35\linewidth]{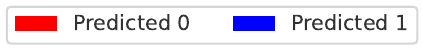}
    \par  
    \begin{subfigure}[b]{0.24\linewidth}
        \includegraphics[width=\linewidth]{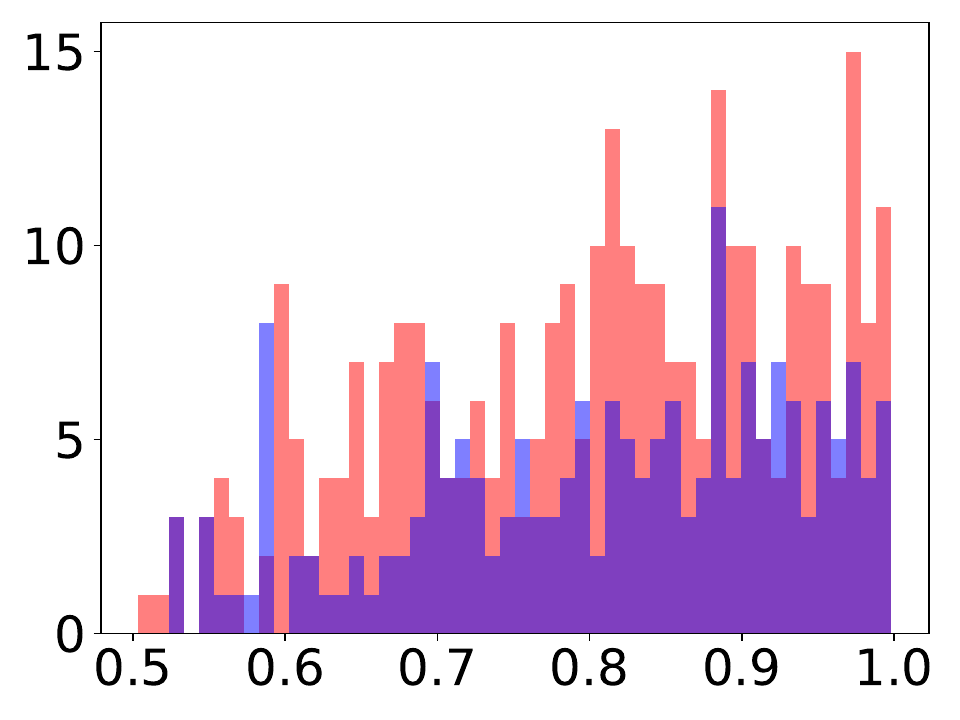}
        \caption{\(L=1\)}
    \end{subfigure}
    \hfill
    \begin{subfigure}[b]{0.24\linewidth}
        \includegraphics[width=\linewidth]{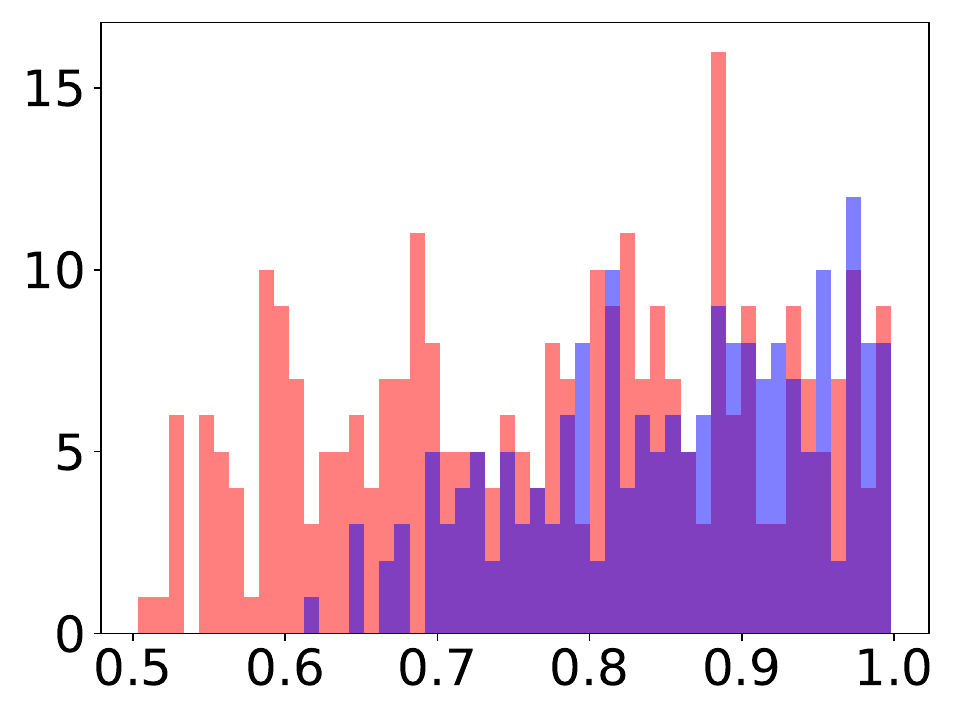}
        \caption{\(L=2\)}
        \label{fig:purity_classification_L2}
    \end{subfigure}
    \hfill
    \begin{subfigure}[b]{0.24\linewidth}
        \includegraphics[width=\linewidth]{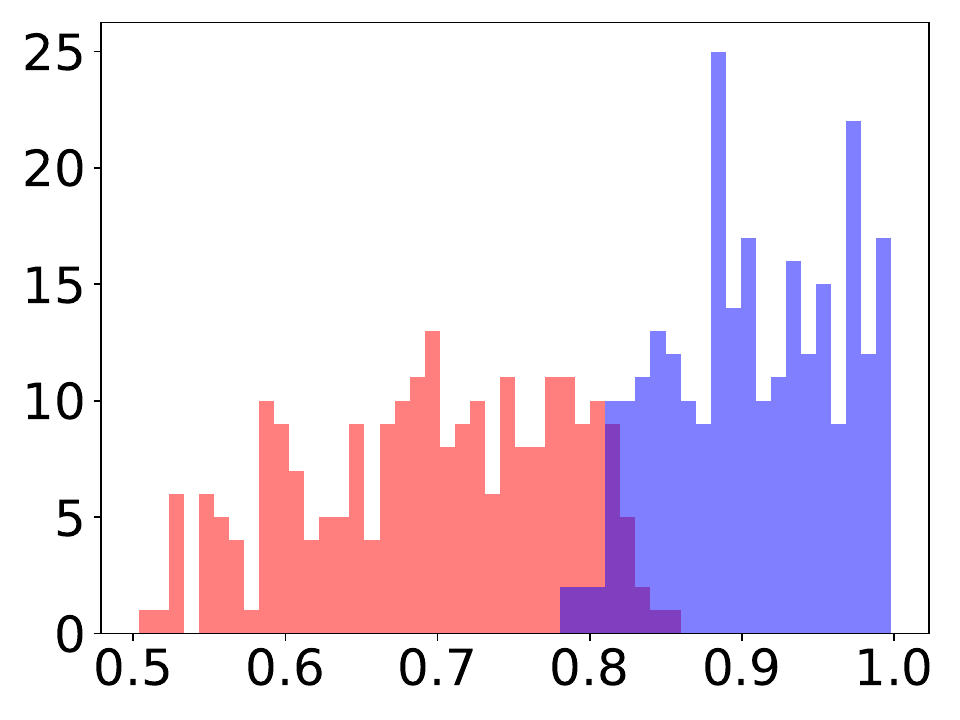}
        \caption{\(L=3\)}
    \end{subfigure}
    \hfill
    \begin{subfigure}[b]{0.24\linewidth}
        \includegraphics[width=\linewidth]{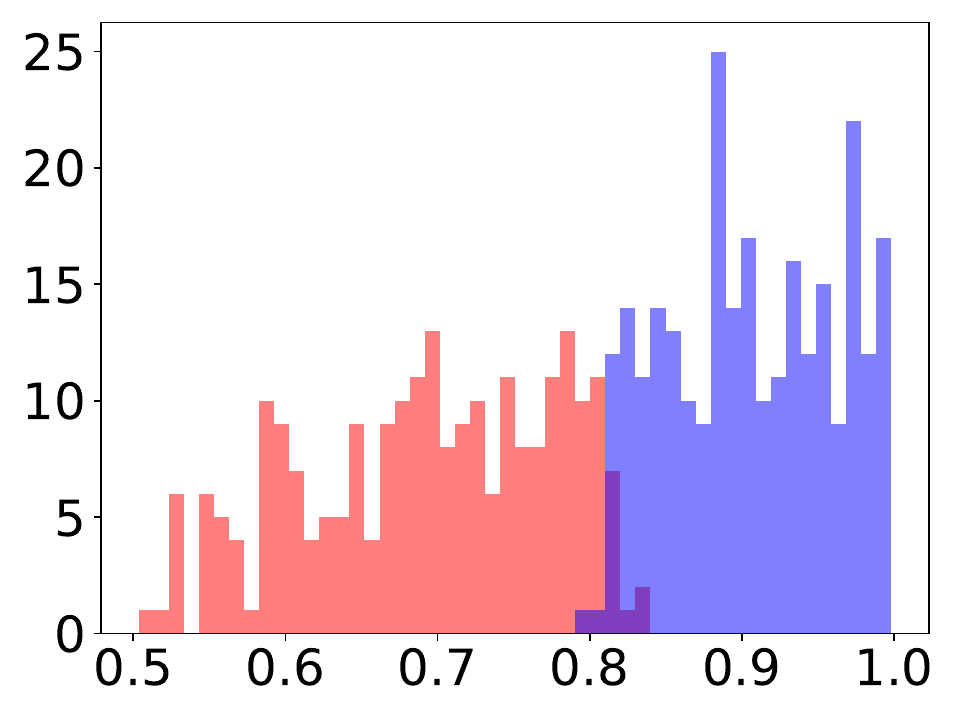}
        \caption{\(L=4\)}
    \end{subfigure}
    \caption{Purity classification histograms for re-uploading depths \(L=1\) to \(L=4\), where the corresponding test accuracies are 0.51, 0.61, 0.95, and 0.98, respectively. The \(x\)-axis shows purity and the \(y\)-axis the sample count. Blue bars indicate predictions of class 1 and red bars predictions of class 0.}
    \label{fig:purity_classification_histogram}
\end{figure}

In this section, we consider the task of purity classification for single-qubit states. The purity of a density matrix \(\rho\) is defined by \(\Tr(\rho^2)\), satisfying \(1/d\le\Tr(\rho^2)\le1\) \cite{nielsen2010quantum}. We sample single-qubit states by choosing Bloch vectors uniformly from the unit ball in \(\R^3\), and formulate a binary classification problem by asking whether the purity of \(\rho\) exceeds the threshold \(\text{th}=(1+2^{-2/3})/2\approx0.815\), assigning class 1 if it does and class 0 otherwise. The value $(1+2^{-2/3})/2$ ensures balanced classes, i.e., \(P(\Tr(\rho^2)<\text{th})=P(\Tr(\rho^2)\ge\text{th})=0.5\). We generated a training set of 1000 states and a test set of 500 states, and trained the unrestricted re-uploading circuit in Figure~\ref{fig:reupload_general} to perform the classification. Results are shown in Figures~\ref{fig:purity_classification_accuracy_curve} and \ref{fig:purity_classification_histogram}.

The re-uploading model is strictly less expressive than a global \((Ln+1)\)-qubit unitary, even if we restrict our focus to system \(A\). For example, we can implement the \emph{purity test} circuit by allowing an entangling operation between just two copies of \(\rho\) (see Appendix~\ref{section:purity_test_derivation} for details), which is a \((2n+1)\)-qubit unitary. However, the same test cannot be implemented by the re-uploading model with \(L=2\) (which also uses two copies of \(\rho\)), as empirically demonstrated in Figure~\ref{fig:purity_classification_L2} and formally stated in the following observation.
\begin{observation}
\label{observation:purity_impossible}
The re-uploading model with \(L=2\), where the output has the form of Eq.~\eqref{equation:final_function}, cannot implement the purity function.
\end{observation}

\begin{proof}
See Appendix~\ref{section:impossibility_proof}.
\end{proof}

\subsection{Entanglement entropy classification}

\begin{figure}[t]
    \centering
    \includegraphics[width=0.35\linewidth]{purity_classification/legend_box.pdf}
    \par  
    \begin{subfigure}[b]{0.24\linewidth}
        \includegraphics[width=\linewidth]{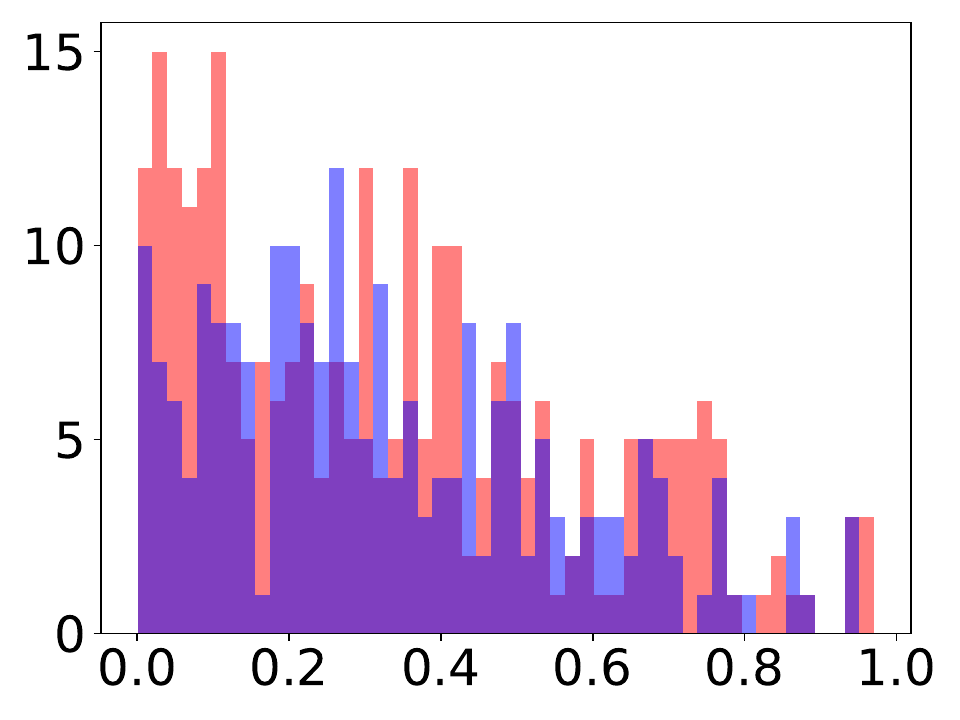}
        \caption{\(L=1\)}
    \end{subfigure}
    \hfill
    \begin{subfigure}[b]{0.24\linewidth}
        \includegraphics[width=\linewidth]{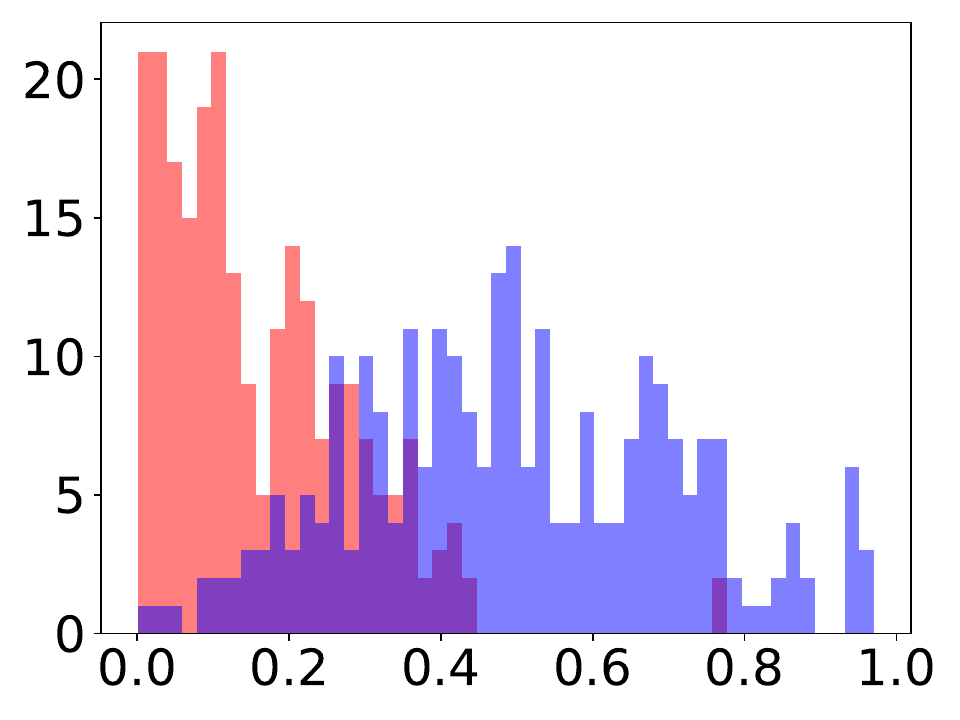}
        \caption{\(L=2\)}
    \end{subfigure}
    \hfill
    \begin{subfigure}[b]{0.24\linewidth}
        \includegraphics[width=\linewidth]{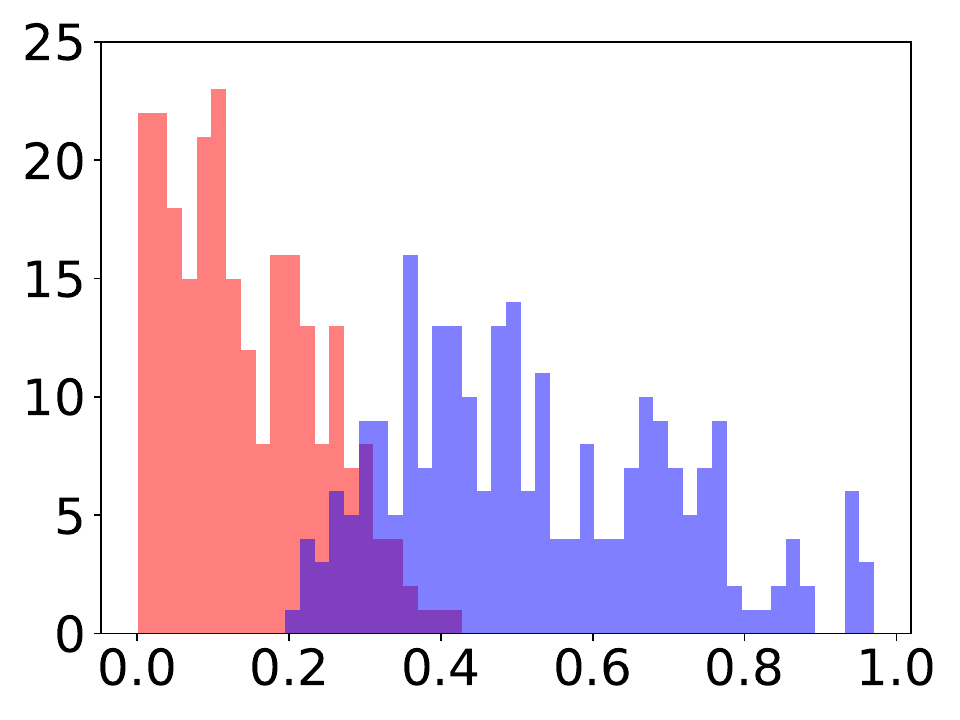}
        \caption{\(L=3\)}
    \end{subfigure}
    \hfill
    \begin{subfigure}[b]{0.24\linewidth}
        \includegraphics[width=\linewidth]{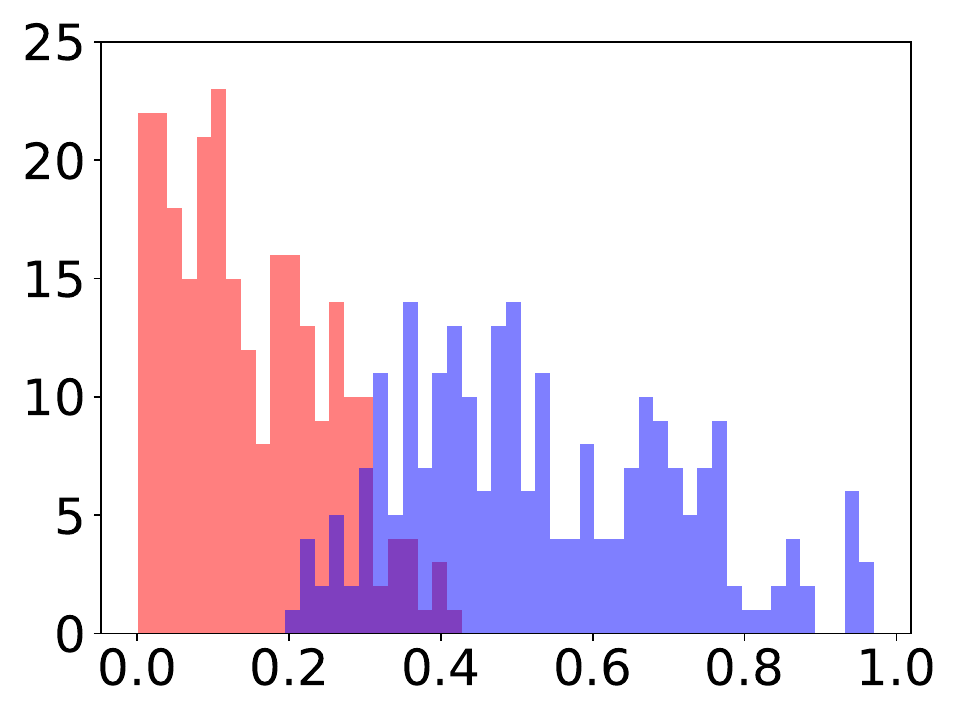}
        \caption{\(L=4\)}
    \end{subfigure}
    \caption{Entanglement entropy classification histograms for re-uploading depths \(L=1\) to \(L=4\), where the corresponding test accuracies are 0.49, 0.84, 0.92, and 0.93, respectively. The \(x\)-axis shows entanglement entropy and the \(y\)-axis the sample count. Blue bars indicate predictions of class 1 and red bars predictions of class 0.}
    \label{fig:entropy_classification_histogram}
\end{figure}

One important nonlinear function of \(\rho\) is the second-order R\'enyi entanglement entropy
\[
-\log(\Tr(\rho^{\otimes2}S_A)),
\]
where \(A\) is a subsystem of \(\rho\) and the local swap operator \(S_A\) acts as
\[
S_A|i_Aj_{A^c}\rangle|k_Al_{A^c}\rangle = |k_Aj_{A^c}\rangle|i_Al_{A^c}\rangle
\]
for computational basis states \(|i\rangle\), \(|j\rangle\), \(|k\rangle\), and \(|l\rangle\). We sampled 2-qubit pure states according to the Haar measure \cite{mele2024introduction} and chose \(A\) to be the first qubit. Then we assigned label 1 to states with entanglement entropy \(\ge0.3\), which yields approximately balanced classes. We generated a training set of 1000 states and a test set of 500 states, and trained the unrestricted re-uploading circuit in Figure~\ref{fig:reupload_general} to perform the classification. Results are shown in Figures~\ref{fig:entropy_classification_accuracy_curve} and \ref{fig:entropy_classification_histogram}.

\subsection{Classification on the Bloch Sphere}

\begin{figure}
    \centering
    \includegraphics[width=0.3\linewidth]{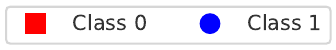}
    \par  
    \begin{subfigure}[b]{0.4\linewidth}
        \includegraphics[width=\linewidth]{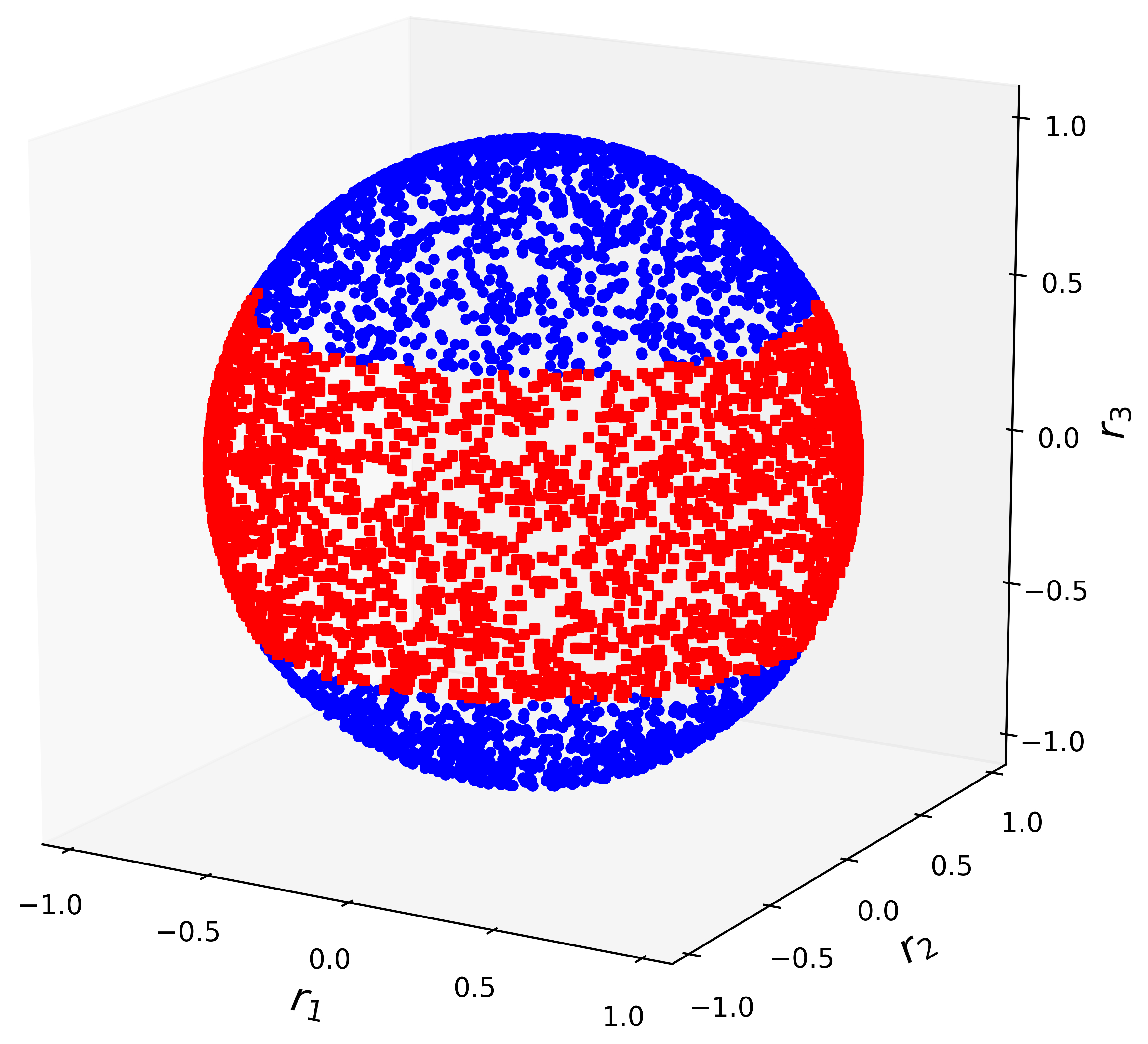}
        \caption{\centering}
        \label{fig:band_dataset}
    \end{subfigure}
    \hspace{1em}
    \begin{subfigure}[b]{0.4\linewidth}
        \includegraphics[width=\linewidth]{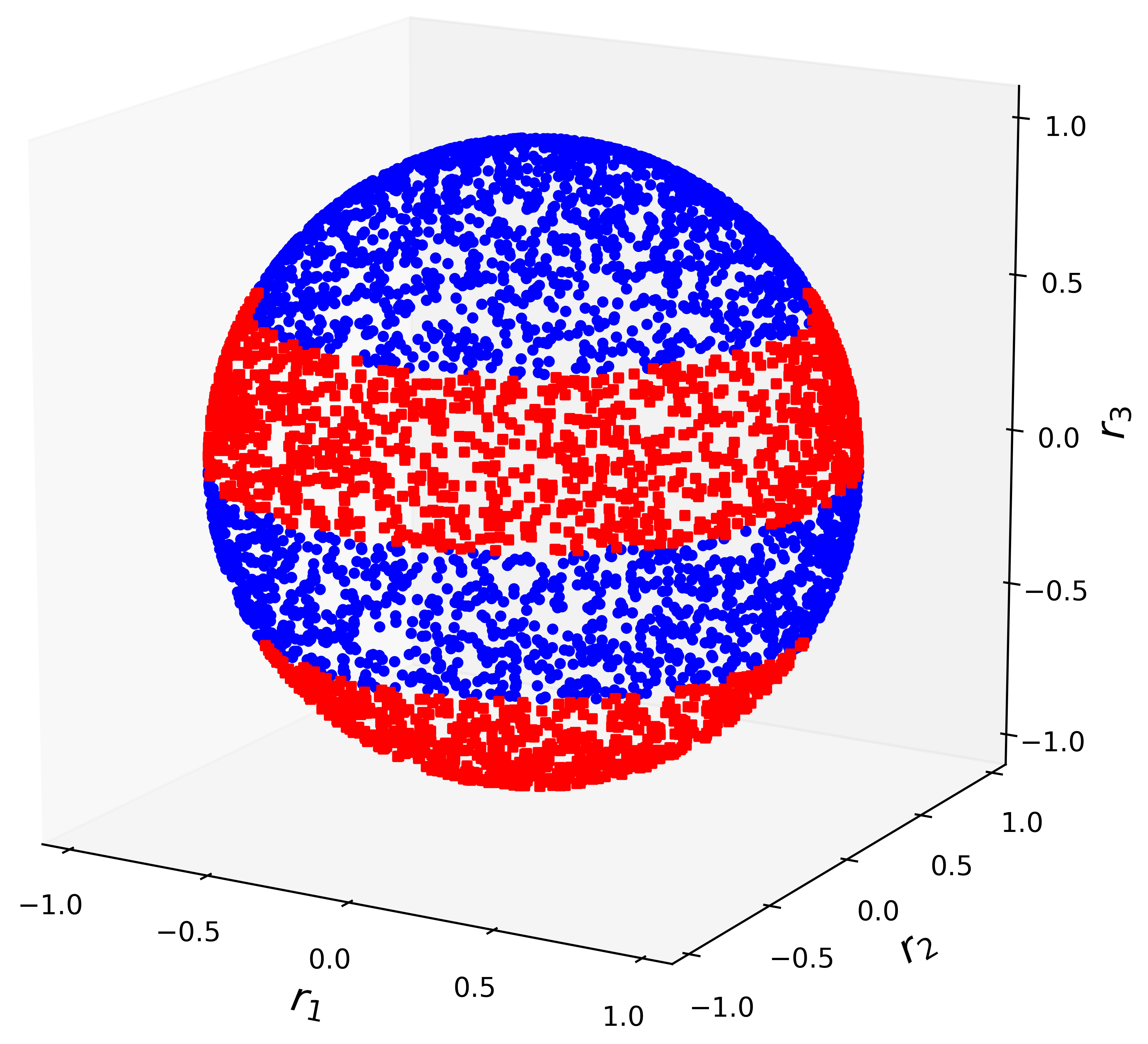}
        \caption{\centering}
        \label{fig:bandband_dataset}
    \end{subfigure}
    \caption{Two classification datasets consisting of unit-norm Bloch vectors \((r_1,r_2,r_3)\). Red squares and blue circles correspond to classes 0 and 1, respectively.}
    \label{fig:band_data_plot}
\end{figure}

\begin{figure}
    \centering
    \includegraphics[width=0.35\linewidth]{purity_classification/legend_box.pdf}
    \par  
    \begin{subfigure}[b]{0.24\linewidth}
        \includegraphics[width=\linewidth]{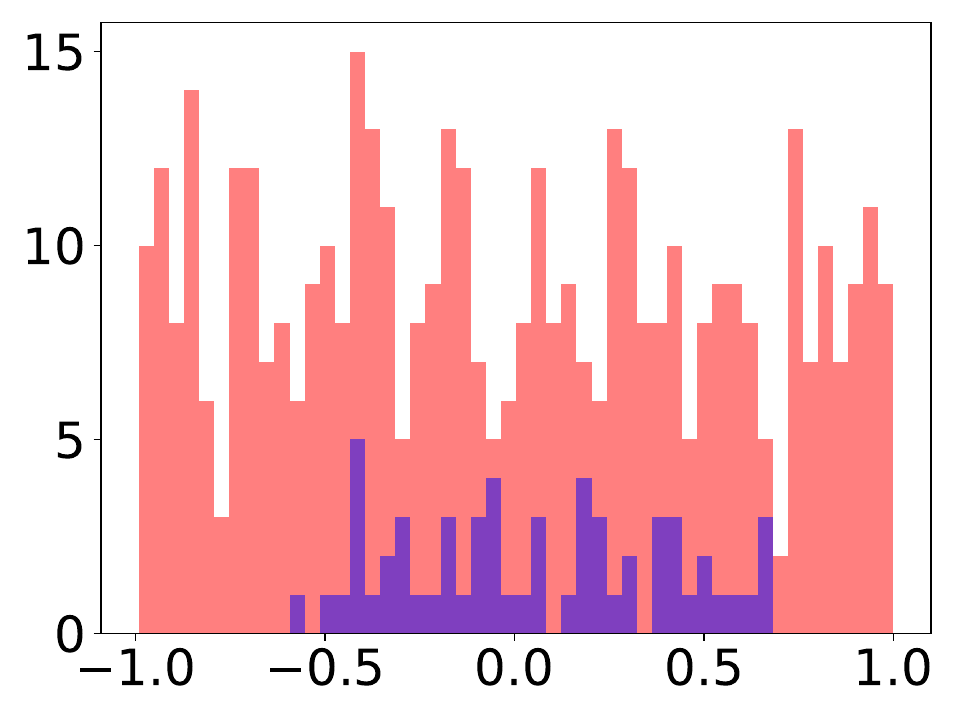}
        \caption{\centering}
    \end{subfigure}
    \hfill
    \begin{subfigure}[b]{0.24\linewidth}
        \includegraphics[width=\linewidth]{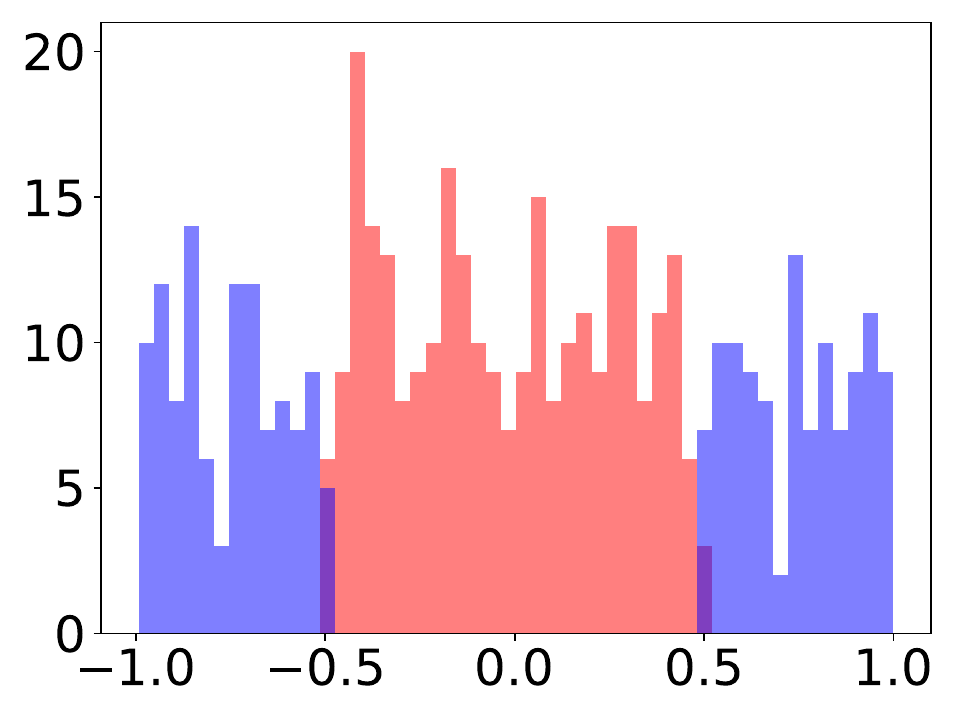}
        \caption{\centering}
    \end{subfigure}
    \hfill
    \begin{subfigure}[b]{0.24\linewidth}
        \includegraphics[width=\linewidth]{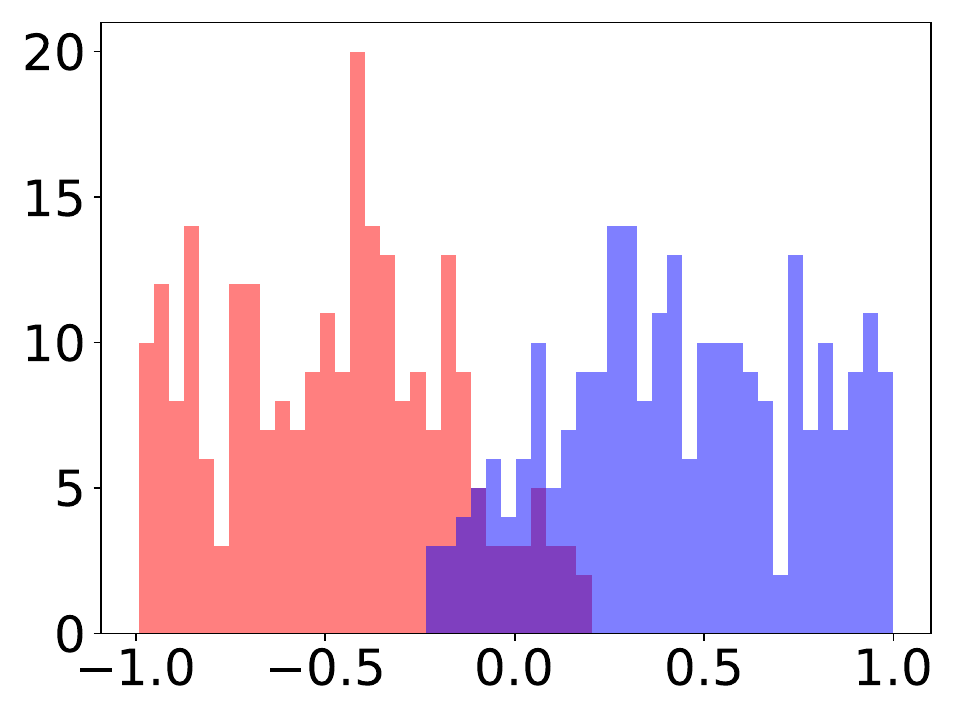}
        \caption{\centering}
    \end{subfigure}
    \hfill
    \begin{subfigure}[b]{0.24\linewidth}
        \includegraphics[width=\linewidth]{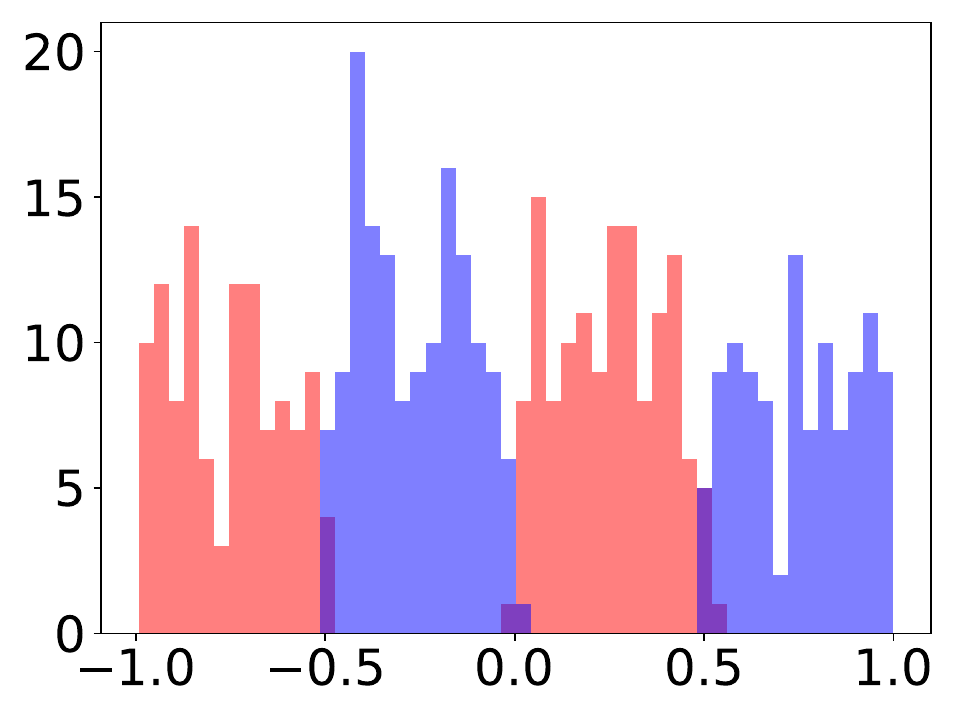}
        \caption{\centering}
    \end{subfigure}
    \caption{Bloch sphere classification histograms for (a) Figure~\ref{fig:band_dataset} with \(L=1\), (b) Figure~\ref{fig:band_dataset} with \(L=2\), (c) Figure~\ref{fig:bandband_dataset} with \(L=2\), and (d) Figure~\ref{fig:bandband_dataset} with \(L=3\), where the corresponding test accuracies are 0.47, 0.99, 0.53, and 0.99, respectively. The \(x\)-axis shows \(r_3\) and the \(y\)-axis the sample count. Blue bars indicate predictions of class 1 and red bars predictions of class 0.}
    \label{fig:band_classification_histogram}
\end{figure}

In this section, we test the circuit's ability to learn a different nonlinear decision boundary by classifying pure single-qubit states on the Bloch sphere. We sampled Bloch vectors uniformly from the unit sphere and assigned label 1 to states satisfying either ``\(|r_3|\ge0.5\)'' (see Figure~\ref{fig:band_dataset}) or ``\(r_3\ge0.5\text{ or }-0.5\le r_3<0\)'' (see Figure~\ref{fig:bandband_dataset}). Each labeling rule yields balanced classes. We generated a training set of 1000 states and a test set of 500 states, and trained the unrestricted re-uploading circuit in Figure~\ref{fig:reupload_general} to perform the classification. As shown in Figure~\ref{fig:band_classification_histogram}, an accuracy of 0.99 is achieved for (i) Figure~\ref{fig:band_dataset} with \(L=2\) and (ii) Figure~\ref{fig:bandband_dataset} with \(L=3\). This is consistent with the fact that (i) two re-uploading layers suffice to implement \(f(r_1,r_2,r_3)=r_3^2\) (with threshold 0.25 for prediction) and (ii) three re-uploading layers suffice to implement \(f(r_1,r_2,r_3)=r_3(r_3-0.5)(r_3+0.5)\) (with threshold 0 for prediction).

\section{Discussion and outlook}

There exists a fundamental trade-off between expressivity and resource overhead in quantum data re-uploading. On one end, performing successive uploads of a single copy of \(\rho\) interleaved with an \((n+1)\)-qubit unitary reduces qubit count. On the other end, uploading \(L\) copies in parallel and applying an \((Ln+1)\)-qubit unitary maximizes expressivity at the cost of hardware scalability. Characterizing the optimal balance between these extremes remains a challenge.

\hl{From a trainability standpoint, the re-uploading ansatz may offer a relative practical advantage over parallel multicopy ansatzes:} by keeping the register size small, it might exhibit less severe barren-plateau issues known to worsen with a larger number of qubits \cite{mcclean2018barren}. Nonetheless, a detailed ansatz-dependent landscape analysis is essential to guide implementation in near-term devices.

While our universality proof focuses on a single-qubit signal register, extending the framework to the multi-qubit case could enhance model capabilities. A systematic study of how expressivity scales with the size of the signal register, and whether this yields more efficient approximations for particular function classes, is a compelling direction for future work. Moreover, our expressivity analysis was confined to a narrow family of ansatz-agnostic operations involving alternating single-qubit rotations and controlled unitaries, together with specific constructions for polynomial realization. A comprehensive investigation of more general and practical operations remains largely open.

Finally, identifying concrete learning tasks where re-uploading outperforms traditional tomography-plus-post-processing pipelines will be key to demonstrating practical advantage. This entails designing optimal ansatzes for re-uploading circuits tailored to specific quantum datasets.

\section*{Author contributions}

H.C. led the project, conceived the main idea, developed the main proofs, and wrote the initial draft of the manuscript. D.K.P. established the connection to DQC1, and assisted with physical interpretation and experiments. J.L. helped structure the manuscript, reviewed the text, and assisted in identifying and correcting mathematical errors.

\begin{acknowledgments}
This work is in part supported by the National Research Foundation of Korea (NRF, RS-2024-00451435 (10\%), RS-2024-00413957 (10\%), RS-2023-NR119931 (10\%), RS-2025-02309510 (10\%)), Institute of Information \& Communications Technology Planning \& Evaluation (IITP, RS-2021-II212068 (10\%), RS-2025-02305453 (10\%), RS-2025-02273157 (10\%), RS-2025-25442149 (10\%), RS-2021-II211343 (10\%), No. 2019-0-00003 (10\%); Research and Development of Core Technologies for Programming, Running, Implementing and Validating of Fault-Tolerant Quantum Computing System) grant funded by the Ministry of Science and ICT (MSIT), Institute of New Media and Communications (INMAC), and the BK21 FOUR program of the Education, Artificial Intelligence Graduate School Program (Seoul National University), and Research Program for Future ICT Pioneers, Seoul National University in 2025.
\end{acknowledgments}

\bibliographystyle{quantum}
\bibliography{references}

\appendix

\section{Derivation of Eq.~\eqref{equation:evolution_formula}}
\label{section:evolution_formula_derivation}

Note that
\begin{align*}
\sum_{\beta=1}^{d/2} \braauto{\psi_\beta^{(\alpha)}} W_{\alpha^\prime} \ketauto{\psi_\beta^{(\alpha)}} & = \Tr\!\left( \sum_{\beta=1}^{d/2} \ketbraauto{\psi_\beta^{(\alpha)}} W_{\alpha^\prime} \right) = \Tr(W_\alpha^+ W_{\alpha^\prime}),\\
\sum_{\gamma=1}^{d/2} \braauto{\phi_\gamma^{(\alpha)}} W_{\alpha^\prime} \ketauto{\phi_\gamma^{(\alpha)}} & = \Tr\!\left( \sum_{\gamma=1}^{d/2} \ketbraauto{\phi_\gamma^{(\alpha)}} W_{\alpha^\prime} \right) = \Tr(W_\alpha^- W_{\alpha^\prime}),
\end{align*}
and
\begin{align*}
\Tr(W_\alpha^+ W_{\alpha^\prime}) + \Tr(W_\alpha^- W_{\alpha^\prime}) & = \Tr(W_{\alpha^\prime}) = 0,\\
\Tr(W_\alpha^+ W_{\alpha^\prime}) - \Tr(W_\alpha^- W_{\alpha^\prime}) & = \Tr(W_\alpha W_{\alpha^\prime}) = d\cdot\delta_{\alpha,\alpha^\prime}.
\end{align*}
Then we have
\begin{align*}
\tau^{(l)} & = \sum_{\beta=1}^{d/2} K_\beta^{(\alpha+)} \left(\Tilde{\tau}^{(l)}\otimes\rho\right) K_\beta^{(\alpha+)\dagger} + \sum_{\gamma=1}^{d/2} K_\gamma^{(\alpha-)} \left(\Tilde{\tau}^{(l)}\otimes\rho\right) K_\gamma^{(\alpha-)\dagger}\\
& = \frac{1}{2d} \left( \I_2 + \Tilde{r}_1^{(l)}X + \Tilde{r}_2^{(l)}Y + \Tilde{r}_3^{(l)}Z \right) \sum_{\beta=1}^{d/2} \left( 1 + \sum_{\alpha^\prime=1}^{d^2-1}\lambda_{\alpha^\prime} \braauto{\psi_\beta^{(\alpha)}} W_{\alpha^\prime} \ketauto{\psi_\beta^{(\alpha)}} \right)\\
& + \frac{1}{2d} \left( \I_2 + \Tilde{r}_1^{(l)}X - \Tilde{r}_2^{(l)}Y - \Tilde{r}_3^{(l)}Z \right) \sum_{\gamma=1}^{d/2} \left( 1 + \sum_{\alpha^\prime=1}^{d^2-1}\lambda_{\alpha^\prime} \braauto{\phi_\gamma^{(\alpha)}} W_{\alpha^\prime} \ketauto{\phi_\gamma^{(\alpha)}} \right)\\
& = \frac{1}{2d} \left( \I_2 + \Tilde{r}_1^{(l)}X \right) \left( d + \sum_{\alpha^\prime=1}^{d^2-1}\lambda_{\alpha^\prime} \left[ \Tr(W_\alpha^+ W_{\alpha^\prime}) + \Tr(W_\alpha^- W_{\alpha^\prime}) \right] \right)\\
& + \frac{1}{2d} \left( \Tilde{r}_2^{(l)}Y + \Tilde{r}_3^{(l)}Z \right) \sum_{\alpha^\prime=1}^{d^2-1}\lambda_{\alpha^\prime} \left[ \Tr(W_\alpha^+ W_{\alpha^\prime}) - \Tr(W_\alpha^- W_{\alpha^\prime}) \right]\\
& = \frac{1}{2} \left( \I_2 + \Tilde{r}_1^{(l)}X \right) + \frac{1}{2d} \left( \Tilde{r}_2^{(l)}Y + \Tilde{r}_3^{(l)}Z \right) \sum_{\alpha^\prime=1}^{d^2-1}\lambda_{\alpha^\prime} d\cdot\delta_{\alpha,\alpha^\prime}\\
& = \frac{1}{2} \left( \I_2 + \Tilde{r}_1^{(l)}X \right) + \frac{1}{2} \left( \Tilde{r}_2^{(l)}Y + \Tilde{r}_3^{(l)}Z \right) \lambda_\alpha\\
& = \frac{1}{2} \left( \I_2 + \Tilde{r}_1^{(l)}X  + [\lambda_\alpha\Tilde{r}_2^{(l)}]Y + [\lambda_\alpha\Tilde{r}_3^{(l)}]Z \right).
\end{align*}

\section{Derivation of Eq.~\eqref{equation:weak_collision_cos_sin_rescue}}
\label{sec:exp-expand}

\begin{align*}
U_{\alpha,i}(\delta)
&=
\sum_{q=0}^{\infty}\frac{(-i\delta)^q}{q!}H_{\alpha,i}^q \nonumber\\
&=
\sum_{m=0}^{\infty}\frac{(-i\delta)^{2m}}{(2m)!}H_{\alpha,i}^{2m}
+
\sum_{m=0}^{\infty}\frac{(-i\delta)^{2m+1}}{(2m+1)!}H_{\alpha,i}^{2m+1} \nonumber\\
&=
\left(\sum_{m=0}^{\infty}\frac{(-1)^m\delta^{2m}}{(2m)!}\right)(\I_2)_A\otimes(\I_d)_B
-i\left(\sum_{m=0}^{\infty}\frac{(-1)^m\delta^{2m+1}}{(2m+1)!}\right)
\sigma_A^{(i)}\otimes(W_\alpha)_B \nonumber\\
&=
\cos\delta\, (\I_2)_A\otimes(\I_d)_B
-i\sin\delta\,\sigma_A^{(i)}\otimes(W_\alpha)_B
\end{align*}

\begin{highlightblock}
\section{Proof of Theorem~\ref{theorem:quantum_data_reuploading_universality}}
\label{section:formal_universality}

\begin{proof}
First, the generalized Bloch coordinate map is faithful.  Indeed, \(\rho\mapsto\boldsymbol\lambda(\rho)\) is an affine homeomorphism
from \(\mathcal{D}(\mathcal{H}_d)\) onto the compact physical Bloch body
\(\Lambda_d\).  Hence the target induces a continuous function
\(f_\Lambda:\Lambda_d\to\R\), defined by
\(f_\Lambda(\boldsymbol\lambda(\rho))=F(\rho)\).  By the Tietze extension
theorem, extend \(f_\Lambda\) to a continuous function
\(f:[-1,1]^m\to\R\).  Choose constants \(a>0\) and \(b\in\R\) such that
\[
    g(\boldsymbol\lambda)=\frac{f(\boldsymbol\lambda)-b}{a}
    \in[-1,1]
    \qquad \text{for all }\boldsymbol\lambda\in[-1,1]^m .
\]
In particular, one may choose \(a\) to be the right-hand side of Eq.~\eqref{eq:readout-obs-upper-bound}.
After the affine change of variables
\(\mathbf{x}=(\boldsymbol\lambda+\mathbf{1})/2\in[0,1]^m\), define
\(g^\sharp(\mathbf{x})=g(2\mathbf{x}-\mathbf{1})\).

We now invoke the classical re-uploading universality. The underlying single-qubit
classical data re-uploading universality in \(\norm{\cdot}_\infty\) is established in Ref.~\cite{perez2025universal}.  It implies that, for any \(\eta_{\rm cl}>0\), there exist a finite single-qubit classical re-uploading circuit $\mathcal{C}_{\boldsymbol\theta}$ and an observable
\(O_{\rm cl}\in\operatorname{Herm}(\mathcal{H}_2)\) with
\(\norm{O_{\rm cl}}_\infty\le1\) such that
\begin{equation}
\label{equation:classical_expectation_approximation}
    \sup_{\mathbf{x}\in[0,1]^m}
    \left|
        \Tr\!\left(O_{\rm cl}\,
        \mathcal{C}_{\boldsymbol\theta}(2\mathbf{x}-\mathbf{1})(|0\rangle\langle0|)
        \right)
        -g^\sharp(\mathbf{x})
    \right|<\eta_{\rm cl} .
\end{equation}
The circuit uses single-qubit rotations whose angles are
affine functions of the coordinates of \(\mathbf{x}\).  Since
\(x_\alpha=(\lambda_\alpha+1)/2\), each factor depending on
\(x_\alpha\) can be written as an input-independent single-qubit rotation
composed with a data-dependent rotation of the form
\(\exp(-it\lambda_\alpha\sigma^{(i)})\), after possibly changing the real
parameter \(t\) and absorbing global phases.  Therefore the circuit can
be written in the form of Eq.~\eqref{equation:classical_reuploading_circuit_rescue}.

It remains to show that each data-dependent classical rotation in
\(\mathcal{C}_{\boldsymbol\theta}\) is implementable by quantum uploads of
\(\rho\).  Fix \(\alpha\), \(i\), and \(\delta\) with \(|\delta|\le1\).  The
weak collision channel derived in
Eq.~\eqref{equation:exact_weak_collision_channel_rescue} is
\[
    \Phi_{\alpha,i}^{(\delta)}(\tau)
    =\cos^2\delta\,\tau+
    \sin^2\delta\,\sigma^{(i)}\tau\sigma^{(i)}
    -i\lambda_\alpha(\rho)\sin\delta\cos\delta\,[\sigma^{(i)},\tau],
\]
whereas the ideal classical rotation channel in
Eq.~\eqref{equation:exact_ideal_rotation_channel_rescue} is
\[
    \mathcal{R}_{\alpha,i}^{(\delta)}(\tau)
    =\cos^2(\delta\lambda_\alpha)\tau+
    \sin^2(\delta\lambda_\alpha)\sigma^{(i)}\tau\sigma^{(i)}
    -i\sin(\delta\lambda_\alpha)\cos(\delta\lambda_\alpha)
    [\sigma^{(i)},\tau].
\]
Using \(|\lambda_\alpha|\le1\), \(\norm{\tau}_1=1\),
\(\norm{\sigma^{(i)}\tau\sigma^{(i)}}_1=1\), and
\(\norm{[\sigma^{(i)},\tau]}_1\le2\), we obtain the uniform single-step
bound
\begin{equation}
\label{equation:formal_single_step_bound}
    \sup_{\rho\in\mathcal{D}(\mathcal{H}_d)}
    \sup_{\tau\in\mathcal{D}(\mathcal{H}_2)}
    \norm{\Phi_{\alpha,i}^{(\delta)}(\tau)
    -\mathcal{R}_{\alpha,i}^{(\delta)}(\tau)}_1
    \le 6\delta^2 .
\end{equation}

Now choose an integer \(N\ge |t|\) and set \(\delta=t/N\).  Since trace
norm is contractive under completely positive trace-preserving maps,
and since the ideal rotations satisfy
\((\mathcal{R}_{\alpha,i}^{(t/N)})^N=\mathcal{R}_{\alpha,i}^{(t)}\), a
standard telescoping estimate gives
\[
    \sup_{\rho,\tau}
    \norm{(\Phi_{\alpha,i}^{(t/N)})^N(\tau)-
    \mathcal{R}_{\alpha,i}^{(t)}(\tau)}_1
    \le
    N\cdot 6(t/N)^2
    =\frac{6t^2}{N} .
\]
Applying the same argument over the \(M\) data-dependent
rotations in \(\mathcal{C}_{\boldsymbol\theta}\), while noting that the
input-independent channels \(\mathcal{V}^{(j)}\) are isometries with
respect to the trace norm, gives Eq.~\eqref{equation:formal_simulation_bound}.

Finally choose \(\eta_{\rm cl}=\varepsilon/(2a)\) in
Eq.~\eqref{equation:classical_expectation_approximation} and choose the
integers \(N_1,\ldots,N_M\) such that
\[
    6\sum_{j=1}^M \frac{t_j^2}{N_j}<\frac{\varepsilon}{2a}.
\]
Let \(\tau_A^{\rm out}(\rho)=
\widehat{\mathcal{C}}_{\boldsymbol\theta,\mathbf{N}}^{\rho}(|0\rangle\langle0|)\),
set \(O_{\rm out}=aO_{\rm cl}\), and set \(b_{\rm out}=b\).  Then, for all
\(\rho\in\mathcal{D}(\mathcal{H}_d)\),
\begin{align*}
\left|
    b_{\rm out}+\Tr\!\left(O_{\rm out}\tau_A^{\rm out}(\rho)\right)-F(\rho)
\right| & \le
    a\left|
        \Tr\!\left(O_{\rm cl}\tau_A^{\rm out}(\rho)\right)-
        \Tr\!\left(O_{\rm cl}\mathcal{C}_{\boldsymbol\theta}(\boldsymbol\lambda(\rho))(|0\rangle\langle0|)\right)
    \right| \\
&\quad+
    a\left|
        \Tr\!\left(O_{\rm cl}\mathcal{C}_{\boldsymbol\theta}(\boldsymbol\lambda(\rho))(|0\rangle\langle0|)\right)
        -g(\boldsymbol\lambda(\rho))
    \right| \\
& < a\cdot\frac{\varepsilon}{2a}+a\cdot\frac{\varepsilon}{2a}
    =\varepsilon .
\end{align*}
Taking the supremum over \(\rho\) proves the claimed uniform
approximation guarantee.  The readout is an affine combination of
single-qubit Pauli operators because every Hermitian
\(O_{\rm out}\in\operatorname{Herm}(\mathcal{H}_2)\) is a real linear combination
of \(\I_2,X,Y,Z\), and the identity contribution is absorbed into
\(b_{\rm out}\).
\end{proof}
\end{highlightblock}

\section{Details of the purity test}
\label{section:purity_test_derivation}

\begin{figure}[h]
    \centering
    \includegraphics[width=0.5\linewidth]{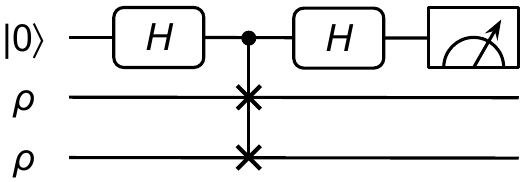}
    \caption{The swap test circuit \cite{barenco1997stabilization, buhrman2001quantum}, typically used to estimate inner products between pure states, also serves as a tool for estimating the purity of a quantum state \(\rho\).}
    \label{fig:swap_test}
\end{figure}

Figure~\ref{fig:swap_test} shows that applying the swap test to two copies of \(\rho\) enables evaluation of the purity of \(\rho\). In more detail, write
\[
\rho = \sum_i \mu_i \ketbraauto{\psi_i},
\]
where \(\mu_i\ge0\) and \(\{|\psi_i\rangle\}_i\) is an orthonormal basis. Then
\[
\rho^{\otimes2} = \sum_{i,j}\mu_i\mu_j\ketbraauto{\psi_i}\otimes\ketbraauto{\psi_j}
\]
and
\begin{align*}
P\left(\text{First qubit}=0 \mid \rho^{\otimes2}\right) & = \sum_{i,j}\mu_i\mu_j P\left(\text{First qubit}=0 \mid \ket{\psi_i}\otimes\ket{\psi_j}\right)\\
& = \sum_{i,j}\mu_i\mu_j \frac{1}{2}(1+\delta_{i,j})\\
& = \frac{1}{2}\sum_i\mu_i^2 + \frac{1}{2}\left(\sum_i\mu_i\right)^2\\
& = \frac{1}{2}+\frac{1}{2}\Tr(\rho^2).
\end{align*}
The expectation value \(\langle Z\rangle\) on system \(A\) is equal to the purity \(\Tr(\rho^2)\).

\section{Proof of Observation~\ref{observation:purity_impossible}}
\label{section:impossibility_proof}

\begin{proof}
We show that no collection \((U,V,O)\) exists such that \(U,V\in\textnormal{SU}(4)\), \(O\in\operatorname{Herm}(\mathcal{H}_2)\), and \(\langle O_A\rangle_{\sigma_A}=\Tr(\rho^2)\), where
\[
\sigma_A=\Tr_{B_2}\!\left(V_{AB_2}\left(\Tr_{B_1}\!\left(U_{AB_1}(|0\rangle\langle0|_A\otimes\rho_{B_1})U_{AB_1}^\dagger\right)\otimes\rho_{B_2}\right)V_{AB_2}^\dagger\right).
\]
\hl{The proof compares two facts about the effective two-copy observable: the re-uploading form forces a singular Pauli correlation matrix, whereas purity forces a nonsingular one.} Write
\[
K_k = ((\I_2)_A\otimes\braauto{k_{B_1}}) U_{AB_1} (\ketauto{0_A}\otimes(\I_2)_{B_1}) \quad (k=0,1).
\]
After the first upload, the state of system \(A\) is given by
\[
\sum_{k=0}^1 K_k \rho_{B_1} K_k^\dagger.
\]
For the second upload set
\[
J_k = V_{AB_2} (K_k \otimes (\I_2)_{B_2}), \quad \Tilde{O}_{AB_2} = O_A \otimes (\I_2)_{B_2}.
\]
Then
\begin{align*}
\langle O_A\rangle_{\sigma_A} & = \Tr_{AB_2} \!\left( \left[ \sum_{k=0}^1 J_k (\rho_{B_1}\otimes\rho_{B_2}) J_k^\dagger \right] \Tilde{O}_{AB_2} \right)\\
& = \Tr_{B_1B_2} \!\left( (\rho_{B_1}\otimes\rho_{B_2}) \left[ \sum_{k=0}^1 J_k^\dagger \Tilde{O}_{AB_2} J_k \right] \right).
\end{align*}
For \(M\in\operatorname{Herm}(\mathcal{H}_2 \otimes \mathcal{H}_2)\), define \(\text{corr}(M)\in\R^{3\times3}\) such that
\[
\text{corr}(M)_{ij} = \frac{1}{2}\Tr\!\left(M\left(\sigma^{(i)}\otimes\sigma^{(j)}\right)\right).
\]
In what follows, we show that
\[
\det\!\left( \text{corr}\!\left( \sum_{k=0}^1 J_k^\dagger \Tilde{O}_{AB_2} J_k \right) \right) = 0.
\]
For notational simplicity, we omit system subscripts and show that \(\det(T)=0\), where
\begin{gather*}
K_k = (\I_2\otimes\bra{k}) U (\ket{0}\otimes\I_2), \quad J_k = V (K_k \otimes \I_2), \quad \Tilde{O} = O \otimes \I_2,\\
\text{and} \quad T = \text{corr}\!\left( \sum_{k=0}^1 J_k^\dagger \Tilde{O} J_k \right).
\end{gather*}
Let
\[
Q = V^\dagger \Tilde{O} V
\]
and define
\[
\Lambda(\cdot) = \sum_{k=0}^1 K_k(\cdot)K_k^\dagger.
\]
Then
\begin{align*}
T_{ij} & = \frac{1}{2} \Tr\!\left(\sum_{k=0}^1\left(K_k^\dagger\otimes\I_2\right)Q(K_k\otimes\I_2)\left(\sigma^{(i)}\otimes\sigma^{(j)}\right)\right)\\
& = \frac{1}{2} \Tr\!\left(Q\sum_{k=0}^1(K_k\otimes\I_2)\left(\sigma^{(i)}\otimes\sigma^{(j)}\right)\left(K_k^\dagger\otimes\I_2\right)\right)\\
& = \frac{1}{2} \Tr\!\left(Q\left(\sum_{k=0}^1K_k\sigma^{(i)}K_k^\dagger\otimes \sigma^{(j)}\right)\right)\\
& = \frac{1}{2} \Tr\!\left(Q\left(\Lambda\left(\sigma^{(i)}\right)\otimes \sigma^{(j)}\right)\right).
\end{align*}
But note that \(\Lambda(\sigma^{(i)})\) is traceless for \(i=1,2,3\) because
\[
\Tr\!\left(\Lambda\left(\sigma^{(i)}\right)\right) = \Tr\!\left(\sigma^{(i)}\sum_{k=0}^1K_k^\dagger K_k\right) = \Tr\!\left(\sigma^{(i)}\right) = 0.
\]
Let \(C = \text{corr}(Q)\) and define \(\Tilde{C}\in\R^{3\times3}\) such that
\[
\Tilde{C}_{\mu i} = \frac{1}{2}\Tr\!\left(\sigma^{(\mu)}\Lambda\left(\sigma^{(i)}\right)\right), \quad \mu,i\in\{1,2,3\}.
\]
Then \(\Lambda\left(\sigma^{(i)}\right)\) for \(i=1,2,3\) can be expressed as
\begin{align*}
\Lambda\left(\sigma^{(i)}\right) & = \sum_{\mu=1}^3 \frac{1}{2}\Tr\!\left(\sigma^{(\mu)} \Lambda\left(\sigma^{(i)}\right)\right) \sigma^{(\mu)}\\
& = \sum_{\mu=1}^3 \Tilde{C}_{\mu i} \sigma^{(\mu)},
\end{align*}
hence
\begin{align*}
T_{ij} & = \frac{1}{2} \Tr\!\left(Q\left(\sum_{\mu=1}^3 \Tilde{C}_{\mu i} \sigma^{(\mu)}\otimes \sigma^{(j)}\right)\right)\\
& = \sum_{\mu=1}^3 \Tilde{C}_{\mu i} \frac{1}{2} \Tr\!\left(Q\left( \sigma^{(\mu)}\otimes \sigma^{(j)}\right)\right)\\
& = \sum_{\mu=1}^3 \Tilde{C}_{\mu i} C_{\mu j}\\
& = (\Tilde{C}^TC)_{ij}.
\end{align*}
Therefore, it suffices to show that \(\det(C) = \det(\text{corr}(V^\dagger \Tilde{O} V)) = 0\). Note that any \(V\in\text{SU}(4)\) allows the \emph{KAK decomposition}
\[
V = (\mathfrak{A}_1\otimes\mathfrak{B}_1) e^{i\mathbf{k}\cdot\Sigma} (\mathfrak{A}_2\otimes\mathfrak{B}_2),
\]
where \(\mathfrak{A}_1, \mathfrak{A}_2, \mathfrak{B}_1, \mathfrak{B}_2 \in \text{SU}(2)\), \(\mathbf{k}=(k_1,k_2,k_3)\in\R^3\), and
\[
\Sigma = \left(\sigma^{(1)} \otimes \sigma^{(1)}, \sigma^{(2)} \otimes \sigma^{(2)}, \sigma^{(3)} \otimes \sigma^{(3)}\right)
\]
\cite{tucci2005introduction, wierichs2025recursive}. Also, every \(\mathfrak{U}\in\text{SU}(2)\) is associated with \(\mathfrak{R}(\mathfrak{U})\in\text{SO}(3)\) such that
\[
\mathfrak{U}\sigma^{(i)}\mathfrak{U}^\dagger = \sum_{j=1}^3\mathfrak{R}(\mathfrak{U})_{ij}\sigma^{(j)}, \quad i=1,2,3.
\]
Then
\begin{align*}
& \Tr\!\left( V^\dagger \Tilde{O} V \left(\sigma^{(i)}\otimes\sigma^{(j)}\right) \right)\\
= \, & \Tr\!\left( \left(\mathfrak{A}_2^\dagger\otimes\mathfrak{B}_2^\dagger\right) e^{-i\mathbf{k}\cdot\Sigma} \left(\mathfrak{A}_1^\dagger\otimes\mathfrak{B}_1^\dagger\right) (O\otimes\I_2) (\mathfrak{A}_1\otimes\mathfrak{B}_1) e^{i\mathbf{k}\cdot\Sigma} (\mathfrak{A}_2\otimes\mathfrak{B}_2) \left(\sigma^{(i)}\otimes\sigma^{(j)}\right) \right)\\
= \, & \Tr\!\left( e^{-i\mathbf{k}\cdot\Sigma} (O^\prime\otimes\I_2) e^{i\mathbf{k}\cdot\Sigma}\left(\mathfrak{A}_2\sigma^{(i)}\mathfrak{A}_2^\dagger\otimes\mathfrak{B}_2\sigma^{(j)}\mathfrak{B}_2^\dagger\right)\right)\\
= \, & \sum_{p,q=1}^3 \mathfrak{R}(\mathfrak{A}_2)_{ip} \mathfrak{R}(\mathfrak{B}_2)_{jq} \Tr\!\left( e^{-i\mathbf{k}\cdot\Sigma} (O^\prime\otimes\I_2) e^{i\mathbf{k}\cdot\Sigma}\left(\sigma^{(p)}\otimes\sigma^{(q)}\right)\right)\\
= \, & 2\left(\mathfrak{R}(\mathfrak{A}_2) \text{corr}\!\left(e^{-i\mathbf{k}\cdot\Sigma} (O^\prime\otimes\I_2) e^{i\mathbf{k}\cdot\Sigma}\right) \mathfrak{R}(\mathfrak{B}_2)^T\right)_{ij},
\end{align*}
where \(O^\prime=\mathfrak{A}_1^\dagger O \mathfrak{A}_1\).
\hl{The local factors only left- and right-multiply by $\textnormal{SO}(3)$ matrices, so they do not affect whether the determinant vanishes.}
Now it suffices to show that
\[
\det\!\left(\text{corr}\!\left(e^{-i\mathbf{k}\cdot\Sigma} (O^\prime\otimes\I_2) e^{i\mathbf{k}\cdot\Sigma}\right)\right) = 0.
\]
Since \(\left[\sigma^{(i)}\otimes\sigma^{(i)}, \sigma^{(j)}\otimes\sigma^{(j)}\right] = 0\) for any \((i,j)\) and each \(\sigma^{(i)}\otimes\sigma^{(i)}\) squares to the identity, \(e^{i\mathbf{k}\cdot\Sigma}\) can be factorized as
\begin{align*}
e^{i\mathbf{k}\cdot\Sigma} = & \left[\cos k_1 \I_4 + i\sin k_1 \left(\sigma^{(1)}\otimes\sigma^{(1)}\right)\right]\\
& \left[\cos k_2 \I_4 + i\sin k_2 \left(\sigma^{(2)}\otimes\sigma^{(2)}\right)\right] \left[\cos k_3 \I_4 + i\sin k_3 \left(\sigma^{(3)}\otimes\sigma^{(3)}\right)\right].
\end{align*}
A direct calculation yields
\begin{align*}
\text{corr}\!\left(e^{-i\mathbf{k}\cdot\Sigma} \left(\sigma^{(1)}\otimes\I_2\right) e^{i\mathbf{k}\cdot\Sigma}\right) & = 2\left[\cos(2k_2)\sin(2k_3)\mathbf{e}_2\mathbf{e}_3^T - \sin(2k_2)\cos(2k_3)\mathbf{e}_3\mathbf{e}_2^T\right],\\
\text{corr}\!\left(e^{-i\mathbf{k}\cdot\Sigma} \left(\sigma^{(2)}\otimes\I_2\right) e^{i\mathbf{k}\cdot\Sigma}\right) & = -2\left[\cos(2k_1)\sin(2k_3)\mathbf{e}_1\mathbf{e}_3^T - \sin(2k_1)\cos(2k_3)\mathbf{e}_3\mathbf{e}_1^T\right],\\
\text{corr}\!\left(e^{-i\mathbf{k}\cdot\Sigma} \left(\sigma^{(3)}\otimes\I_2\right) e^{i\mathbf{k}\cdot\Sigma}\right) & = 2\left[\cos(2k_1)\sin(2k_2)\mathbf{e}_1\mathbf{e}_2^T - \sin(2k_1)\cos(2k_2)\mathbf{e}_2\mathbf{e}_1^T\right].
\end{align*}
One can verify that any linear combination of these three matrices has determinant zero. Since \(O^\prime\) can be expressed as a linear combination of \(\left\{\sigma^{(i)}\right\}_{i=0}^3\) and \(\text{corr}\Big(e^{-i\mathbf{k}\cdot\Sigma} \Big(\sigma^{(0)}\otimes\I_2\Big) e^{i\mathbf{k}\cdot\Sigma}\Big)\) evaluates to the zero matrix (hence contributing nothing to \(\text{corr}(\cdot)\)), we have
\[
\det\!\left(\text{corr}\!\left(e^{-i\mathbf{k}\cdot\Sigma} (O^\prime\otimes\I_2) e^{i\mathbf{k}\cdot\Sigma}\right)\right) = 0.
\]
Meanwhile, an element-wise expansion of \(\rho\otimes\rho\) reveals that if \(\hat{O}\in\operatorname{Herm}(\mathcal{H}_2 \otimes \mathcal{H}_2)\) satisfies
\[
\Tr((\rho\otimes\rho)\hat{O})=\Tr(\rho^2)
\]
for all \(\rho\in\mathcal{D}(\mathcal{H}_2)\), then it must take the form
\begin{align*}
\hat{O} & = \text{SWAP} + c_1i\left(\mathbf{e}_2\mathbf{e}_3^T - \mathbf{e}_3\mathbf{e}_2^T\right) + c_2\left(\mathbf{e}_2\mathbf{e}_2^T - \mathbf{e}_3\mathbf{e}_3^T\right)\\
& + (c_3+c_4i)\left(\mathbf{e}_1\mathbf{e}_2^T - \mathbf{e}_1\mathbf{e}_3^T\right) + (c_3-c_4i)\left(\mathbf{e}_2\mathbf{e}_1^T - \mathbf{e}_3\mathbf{e}_1^T\right)\\
& + (c_5+c_6i)\left(\mathbf{e}_2\mathbf{e}_4^T - \mathbf{e}_3\mathbf{e}_4^T\right) + (c_5-c_6i)\left(\mathbf{e}_4\mathbf{e}_2^T - \mathbf{e}_4\mathbf{e}_3^T\right)
\end{align*}
for some \(c_1,\dots,c_6\in\R\). This gives
\[
\text{corr}(\hat{O}) = \begin{pmatrix}
1 & c_1 & -(c_3+c_5)\\
-c_1 & 1 & c_4+c_6\\
c_3+c_5 & -(c_4+c_6) & 1
\end{pmatrix},
\]
which has determinant \(1+c_1^2+(c_3+c_5)^2+(c_4+c_6)^2>0\). \hl{Since the same effective observable cannot have both $\det=0$ and $\det>0$, this concludes the proof of Observation~\ref{observation:purity_impossible}.}
\end{proof}

\end{document}